\documentclass{aa}
\usepackage{epsfig}
\usepackage{txfonts}
\bibpunct{(}{)}{;}{a}{}{,}

\newcommand{\target}{\object{KIC\,7668647}}
\newcommand{\kep}{{\em Kepler}}

\newcommand{\porb}{$P_{\rm{orb}}$}
\newcommand{\teff}{\ensuremath{T_{\rm{eff}}}}
\newcommand{\logg}{\ensuremath{\log g}}

\newcommand{\lheh}{\ensuremath{\log \left(N_{\mathrm{He}}/N_{\mathrm{H}}\right)}}
\newcommand{\ua}{\ensuremath{\uparrow}}
\newcommand{\da}{\ensuremath{\downarrow}}

\renewcommand{\textfraction}{0}
\renewcommand{\dbltopfraction}{1}

\begin{document}



\title{ \target: a 14 day beaming sdB+WD binary \\ with a pulsating
  subdwarf\thanks{Based on observations obtained by the
    \kep\ spacecraft, the Nordic Optical Telescope and the William
    Herschel Telescope}}

\author{
       J.~H.~Telting    \inst{1}  
 \and  A.~S.~Baran      \inst{2}
 \and  P.~Nemeth        \inst{3}
 \and  R.~H.~\O stensen \inst{3}
 \and \\
       T.~Kupfer        \inst{4}
 \and  S.~Macfarlane    \inst{4}
 \and  U.~Heber         \inst{5}
 \and  C.~Aerts         \inst{3,4}
 \and  S.~Geier         \inst{6}
}

\institute{
Nordic Optical Telescope, 
Rambla Jos\'e Ana Fern\'andez P\'erez 7, 38711 Bre\~na Baja,
Spain\\     e-mail: {\tt jht@not.iac.es}
\and
Uniwersytet Pedagogiczny w Krakowie, ul. Podchor\c{a}{\.z}ych 2,
30-084 Krak{\'o}w, Poland
\and
Instituut voor Sterrenkunde, KU Leuven, Celestijnenlaan 200D,
B-3001 Leuven, Belgium
\and
Department of Astrophysics, IMAPP, Radboud University Nijmegen, P.O. Box
9010, NL-6500 GL Nijmegen, the Netherlands
\and
Dr. Remeis Sternwarte, Universit\"at  Erlangen-N\"urnberg, 
Sternwartstr.\ 7, 96049 Bamberg, Germany
\and
European Southern Observatory, Karl-Schwarzschild-Str. 2, 85748 Garching,
Germany
}

\date{submitted           ; accepted           }

\titlerunning{KIC\,7668647: a 14 day beaming sdBV+WD binary}
\authorrunning{J.~H.~Telting et al.}

\abstract{ 
  The recently discovered subdwarf B (sdB) pulsator \target\ is one of
  the 18 pulsating sdB stars detected in the \kep\ field.  It features
  a rich $g$-mode frequency spectrum, with a few low-amplitude
  $p$-modes at short periods. This makes it a promising target for a
  seismic study aiming to constrain the internal structure of this
  star, and of sdB stars in general.  ~~~
  We use new ground-based low-resolution spectroscopy, and the
  near-continuous 2.88\,year \kep\ lightcurve, to reveal that
  \target\ consists of a subdwarf B star with an unseen white-dwarf
  companion with an orbital period of 14.2\,d. An orbit with a
  radial-velocity amplitude of 39\,km\,s$^{-1}$ is consistently
  determined from the spectra, from the orbital Doppler beaming seen
  by \kep\ at 163\,ppm, and from measuring the orbital light-travel
  delay of 27\,s by timing of the many pulsations seen in the
  \kep\ lightcurve.  The white dwarf has a minimum mass of
  0.40\,$M_{\sun}$.  ~~~
  We use our high signal-to-noise average spectra to study the
  atmospheric parameters of the sdB star, and find that nitrogen and
  iron have abundances close to solar values, while helium, carbon,
  oxygen and silicon are underabundant relative to the solar mixture.  ~~~
  We use the full \kep\ Q06--Q17 lightcurve to extract 132 significant
  pulsation frequencies. Period-spacing relations and multiplet
  splittings allow us to identify the modal degree $\ell$ for the
  majority of the modes. Using the { $g$-mode} multiplet splittings we
  constrain the internal rotation period at the base of the
  envelope to 46--48\,d as a first seismic result for this star.
  The few $p$-mode splittings may point at a slightly longer rotation
  period further out in the envelope of the star. ~~~
  From mode-visibility considerations we derive that the inclination of
  the rotation axis of the sdB in \target\ must be around $\sim$60$\degr$. ~~~
  Furthermore, we find strong evidence for a few multiplets indicative
  of degree 3$\le$$\ell$$\le$8, which is another novelty in sdB-star
  observations made possible by \kep.
}

\keywords{ stars: subdwarfs -- stars: early-type, binaries,
  oscillations -- stars: variable: subdwarf-B stars -- stars:
  individual: \mbox{KIC\,7668647} }

\maketitle

\section{Introduction}

The hot subdwarf\,B (sdB) stars populate the extension of the
horizontal branch where the hydrogen envelope mass is too low for
hydrogen burning. These core-helium burning stars must have suffered
extensive mass loss close to the tip of the red giant branch in order
to reach this peculiar configuration of a nearly-exposed hot core with
a relatively thin envelope. Binary interactions, either through stable
Roche lobe overflow or common envelope ejection, are likely to be
responsible for the majority of the sdB population \citep[see][for a
  detailed review]{heber09}.

Recent extensive radial-velocity surveys of sdB stars
\citep{copperwheat11,geier11a} find that $\sim$50\%\ of all
sdB stars reside in short-period binary systems with the majority of
companions being white dwarf (WD) stars. Just above a hundred of such
short-period systems have been found, with periods ranging from 0.07
to 27.8\,d. These systems are all characterised by being single-lined
binaries, {\em i.e.} only the sdB stars contribute to the optical
flux, which directly constrains the companion to be either an M-dwarf
or a compact stellar-mass object.

On the other hand, \citet{reed04} find that $\sim$50\%\ of all sdB
stars have IR excess and must have a companion with spectral type no
later than M2, that all should show up as double-lined binaries.
Orbits of such long-period systems, with periods ranging from
$\sim$500 to 1200\,d, are only now being unravelled through
high-resolution spectroscopy \citep[e.g.][]{vos13}.

The period distributions of these different types of binary systems are
important in that they can be used to constrain a number of vaguely
defined parameters used in binary population synthesis models,
including the common-envelope ejection efficiency, the mass and
angular-momentum loss during stable mass transfer, the minimum core
mass for helium ignition, etc.  The seminal binary population study of
\citet{han02,han03} successfully predicts many aspects of the sdB star
population, but the key parameters have a wide range of possible
values. 

A theoretical prediction of the existence of pulsations in sdB stars,
due to an opacity bump associated with iron ionisation in
subphotospheric layers, was made by \citet{charpinet97}. Since both
$p$ and $g$-mode pulsations were discovered in sdB stars
\citep{kilkenny97,green03} the possibilities to derive the internal
structure, and to constrain the lesser known stages of the evolution,
have widened by means of asteroseismology.  Currently the immediate
aims of asteroseismology of sdB stars are to derive the mass of the
He-burning core and the mass of the thin H envelope
\citep[e.g.][]{randall06}, the rotational frequency
\citep[e.g.][]{telting12,reed14} and internal rotation profile
\citep{charpinet08}, the radius, and the composition of the core
\citep[e.g.][]{VanGrootel10,charpinet11b}.  As argued by
\citet{kawaler2009} the detection of the rotation periods and internal
rotation profiles of sdB stars will help in understanding the
evolution of the angular-momentum budget during the sdB formation
process.

Recent observational success has been achieved from splendid light
curves obtained by the CoRoT and \kep\ spacecrafts, delivering largely
uninterrupted time series with unprecedented accuracy for sdB stars.
Overviews of the \kep\ survey stage results for sdB stars were given
by \citet{ostensen10b,ostensen11b}.  From \kep\ data it has become
clear that the $g$-modes in sdB stars can be identified from period
spacings \citep{reed11c}, which together with multiplet
identifications greatly enhance the observational constraints for
seismic studies.  Evidence for stochastic pulsations in the sdB star
\object{KIC\,2991276} was presented by \cite{ostensen14b}.  With \kep\ it has
been possible to resolve the long rotation periods of these pulsating
sdB (sdBV) stars, explaining why in ground-based campaigns often no
multiplets were uncovered.  For the seemingly non-binary sdB star
\object{KIC\,10670103} a rotation period as long as 88\,d was derived
\citep{reed14}, and all binary sdB stars in the \kep\ field with
clearly detected multiplet splittings show rotation periods much
longer than the orbital periods \citep{pablo11,pablo12,telting12}.

In this paper we present our discovery of the binary nature of
\target\ based on low-resolution spectroscopy and the
\kep\ lightcurve.  This object was sampled as part of a spectroscopic
observing campaign to study the binary nature of the sdBVs in the
\kep\ field, for which preliminary results were presented by
\cite{telting14b}.  Parts of the data from this campaign were
presented in case studies on the $p$-mode pulsator \object{KIC\,10139564}
\citep{baran12}, the blue-horizontal-branch pulsator \object{KIC\,1718290}
\citep{ostensen12BHB}, and the sdBV+WD binary \object{KIC\,11558725}
\citep{telting12}. Recent results from this campaign are the binarity
detection (sdB+WD) of the pulsator \object{KIC\,10553698} \citep{ostensen14a},
and non-detection of binarity in the sdB pulsator \object{KIC\,10670103}
\citep{reed14}.

In total 18 pulsating sdB stars were found in the \kep\ field.  From
the characteristic 'reflection'-effect in \kep\ light curves as many
as 4 sdB+dM binaries with pulsating subdwarf components have been
found by \citet{kawaler10b}, \citet{2m1938}, and by \citet{pablo11}.
Very recently, an additional sdBV+dM binary was discovered in the test
\kep\ $K2$-mission field \citep[EQ Psc,][]{jeffery2014}.  Evidence for
planetary companions around the sdB pulsator \object{KIC\,5807616} is
presented by Charpinet et al.\ (2011).  Five more pulsating subdwarfs
in the original \kep\ field were scrutinised for orbital
radial-velocity variations in the literature, and two of these were
found to have WD companions (\object{KIC\,11558725} and
\object{KIC\,10553698}).  Here we present our detection of the third
case of sdB+WD binarity among the sample of pulsating sdB stars in the
\kep\ field.
 
Our target, \target\ or \object{2MASS J19050638+4318310} or
\object{FBS\,1903+432}, has a \kep\ magnitude of 15.40, and the B-band
magnitude is about 15.1, making it the sixth brightest in the
sample. A first description of the spectroscopic properties, and the
pulsational frequency spectrum as found from the 26 day \kep\ survey
dataset, was given by \citet{ostensen11b}, with the source showing
frequencies in the range of 115--345\,$\mu$Hz.  Based on this
relatively short data set already 15 frequencies were identified,
showing the potential of this star for a seismic study.  Subsequently,
\citet{baran11b} derived 20 frequencies in total from the \kep\ survey
data, and the frequencies were identified in terms of
spherical-harmonic degrees by \citet{reed11c}.  As a consequence, the
star was observed by \kep\ from Q06 onwards.  We have analysed data
from the \kep\ quarters Q06 to Q17.  We present the full frequency
spectrum resulting from these near-continuous 2.88 years of
short-cadence \kep\ observations.


\begin{figure*}[t]
\centerline{\psfig{figure=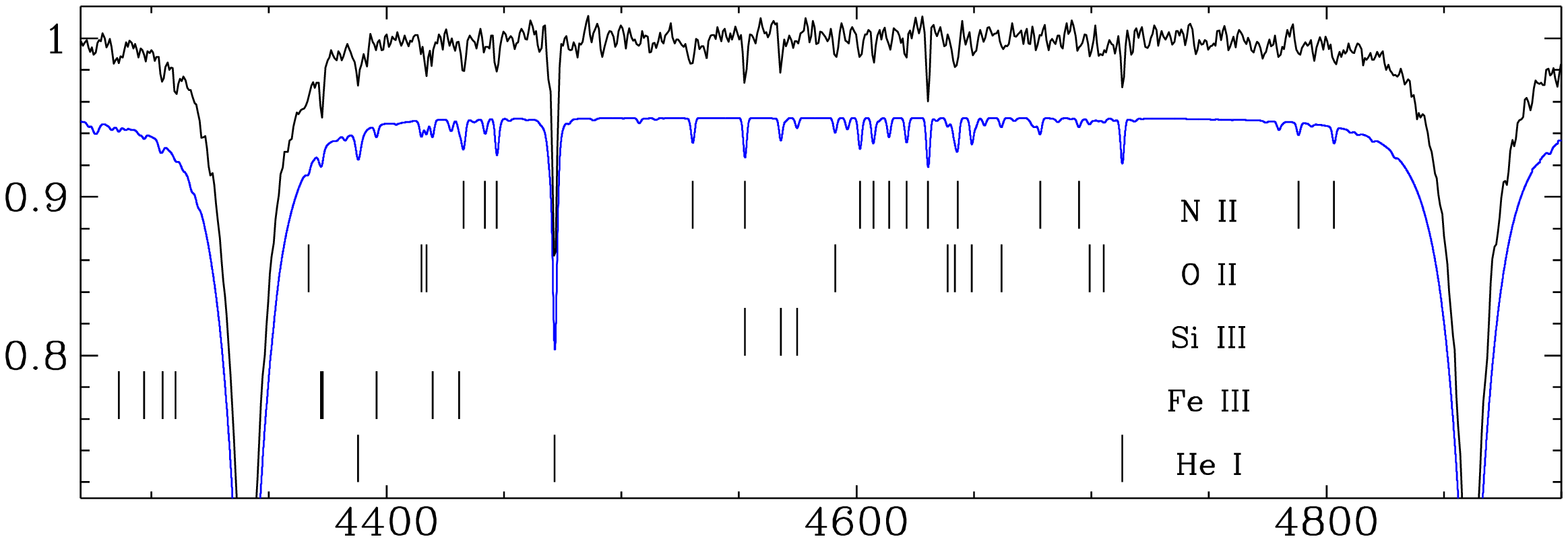,height=11.5cm,angle=-90}}
\centerline{\psfig{figure=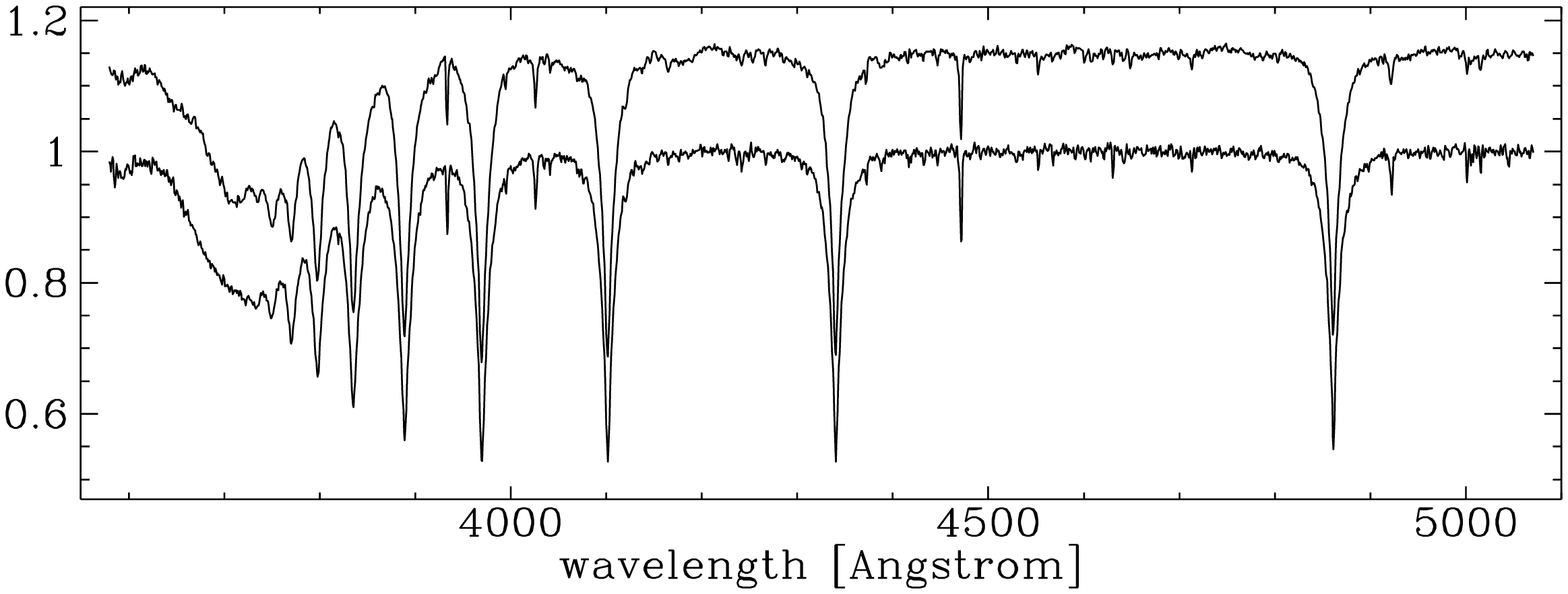,height=11.5cm,angle=-90}}
\caption[]{
  Top:  zoom-in of the mean WHT spectrum with the NLTE fit and line
  identifications, demonstrating some of the stronger lines of heavier
  elements.  The fit is offset for clarity. ~~~~
  Bottom: the mean spectrum from the WHT (bottom) { and NOT, offset in
  flux for clarity.   }}
\label{fig:meanSpectra}
\end{figure*}

From our new spectra of \target\ we solve the orbital radial-velocity
amplitude, and derive a lower limit of the mass of the companion of
the sdB star, which is most likely an unseen white dwarf.  We use the
average spectrum to study the atmospheric parameters in detail.  We
show that the orbital Doppler-boosting (or beaming) signal in the
\kep\ light curve as well as the light travel-time as probed by the
sdB pulsations both lead to independent confirmations of the
spectroscopic orbit solution.  We extract 132 significant pulsational
frequencies, and in the final sections of this paper we discuss
pulsational period spacings and frequency splittings, aiming to
identify the spherical-harmonic degree of the modes and to disclose
the rotation period and inclination angle of the subdwarf in
\target.

\section{Spectroscopic observations}

Over the 2010, 2011, and 2013 observing seasons of the \kep\ field we
obtained altogether 30 spectra of \target.  Low-resolution spectra
(R\,$\approx$\,2000\,--\,2500) have been collected using the 2.56-m Nordic
Optical Telescope with ALFOSC, grism \#16 and a 0.5\,arcsec slit, and
the 4.2-m William Herschel Telescope with ISIS, the R600B grating and
0.9--1.0\,arcsec slit.  Exposure times were 600\,s at WHT, and ranging
between 450\,s and 900\,s at the NOT.  The resulting resolutions based
on the width of arc lines is 1.7\,\AA\ for the WHT setup, and
2.2\,\AA\ for the setup at the NOT.  See Table~\ref{tbl:obslog} for an
observing log.

The data were homogeneously reduced and analysed.  Standard reduction
steps within IRAF include bias subtraction, removal of pixel-to-pixel
sensitivity variations, optimal spectral extraction, and wavelength
calibration based on arc-lamp spectra.  The target spectra and the
mid-exposure times were shifted to the barycentric frame of the solar
system.  The spectra were normalised to place the continuum at unity
by comparing with a model spectrum for a star with similar physical
parameters as we find for the target (see Sect.\ 2.2).  The mean
spectra from each of the telescopes are presented in Fig.\ 1.

\subsection{Radial velocities and orbit solution}

Radial velocities were derived with the FXCOR package in IRAF. We used
the H$\gamma$, H$\delta$, H$\zeta$ and H$\eta$ lines to determine the
radial velocities (RVs), and used the spectral model fit (see next
section) as a template.  For ALFOSC spectra the final RVs were
adjusted for the position of the target in the slit, judged from slit
images taken just before and just after the spectral exposure.  See
Table~\ref{tbl:obslog} for the results, with errors in the radial
velocities as reported by FXCOR.

The errors reported by FXCOR are correct relative to each other, but
may need scaling depending on, amongst other things, the parameter
settings and the validity of the template as a model of the star.  As
our original orbital fit results in a $\chi^2$-value=0.48, the FXCOR
errors seem to be overestimated by about 30\%.  This is consistent
with the RMS found for the orbital fit of about 7.0\,km\,s$^{-1}$.
Therefore we scaled the FXCOR RV errors by a factor of 0.7 to derive
the errors on the orbital parameters below.

Assuming a circular orbit we find an orbital period of 14.1742(42)\,d,
with a radial-velocity amplitude of 38.9(1.9)\,km\,s$^{-1}$ for the
subdwarf. See Table~\ref{tbl:orbit} for the complete parameter
listing.  The radial velocities and the derived solution are shown in
Fig.\ 2.  When fitting an eccentric radial-velocity curve the
eccentricity is fit as $e$=0.056(64).  { We thus
use a circular orbit throughout this paper.}

\begin{figure*}[t]
\centerline{\psfig{figure=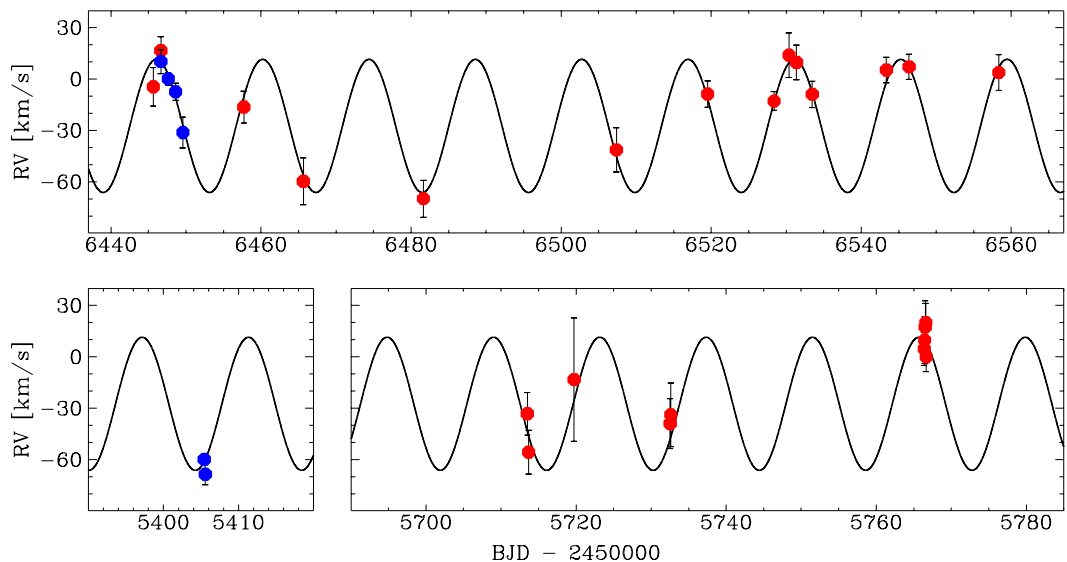,height=14cm,angle=-90}}
\caption[]{Radial-velocity curve from observations with the Nordic
  Optical (red dots) and  William Herschel (blue dots) telescopes.}
\label{fig:radialVelocityCurve}
\end{figure*}

Given the solution presented in Table~\ref{tbl:orbit}, the orbital
radius of the sdB star can be approximated by $a_{\rm sdB} \sin
i$\,=\,10.9(0.5)\,$R_{\sun}$, which corresponds to a light-travel time
of 50.8(2.5) seconds between orbital phases corresponding to closest
and furthest distance to the Sun.


The orbital solution combined with the mass function gives a lower
limit for the mass of the companion of more than 0.40\,$M_{\sun}$, if
one assumes a canonical mass of 0.47\,$M_{\sun}$ 
\citep[see e.g.][]{fontaine12} for the sdB star.  The minimum distance
between the two stars in this binary is then $a \sin
i$\,=\,23\,$R_{\sun}$. As the spectrum does not reveal clear evidence
for light contribution from a companion, it should be either an unseen
compact object, or an unseen M, K or G star.  As the 2MASS
J\,=\,15.82(07) and H\,=\,16.06(20) magnitudes do not indicate a
rising infrared flux that would reveal the presence of a main-sequence
star in this system, we conclude that the unseen companion is most
likely a white dwarf.  
If the inclination of the system is smaller
than $i$\,$\la$\,30$\degr$ then the companion may be a neutron star
or black hole ($i$\,$\la$\,20$\degr$).

\begin{table}[b]
\begin{center}
\caption[]{Orbital solution of \target\ from RV measurements.
$T_{\rm orb,RV}$ is the time at which the subdwarf is closest to
the Sun in its orbit
}
\label{tbl:orbit}
\begin{tabular}{lr@{~}l}
\hline \hline \noalign{\smallskip}
 system velocity [km\,s$^{-1}$]              & $-$27.4  &  (1.3)   \\
 radial-velocity amplitude $K$ [km\,s$^{-1}$]&  38.9    &  (1.9)   \\
 period $P$ [day]                           &  14.1742 &  (0.0042) \\
 $T_{\rm orb,RV}$  [BJD $-$ 2450000]          & 5974.76  &  (0.14)   \\
 reduced $\chi^2$                           &   0.995               \\
 RMS [km\,s$^{-1}$]                          &   7.0                \\
\noalign{\smallskip} \hline
\end{tabular} \end{center} \end{table}

\subsection{Atmospheric parameters}

The spectra were shifted to remove the orbital motion, before being
co-added to obtain high-S/N spectra (S/N\,$>$\,160) with minimal
orbital line broadening, for both observatories.  We derived the
atmospheric parameters of the star from each of these mean spectra,
and list average values of these parameters with errors as
errors-in-mean, in Table~\ref{tbl:specpar}.  For this purpose we used
the fitting procedure of \citet{edelmann03}, with the metal-line
blanketed local thermodynamic equilibrium (LTE) models of solar
composition described in \citet{heber00}.  The errors on the final
adopted values reflect the spread of the individual measurements
rather than the formal errors.  However, as the error-in-mean for the
averaged helium/hydrogen abundance is unrealistically small we use the
smaller of the two individual errors for our final value.  Our final
adopted LTE values are \teff\,=\,27700(300)\,K, \logg\,=\,5.50(3)\,dex
and \lheh\,=\,$-$2.65(3)\,dex, which are quite consistent with the
parameters \teff\,=\,27700\,K, \logg\,=\,5.45\,dex,
\lheh\,=\,$-$2.5\,dex found from an initial NOT survey spectrum in
\citet{ostensen11b}.  Systematic errors related to model physics are
typically of the order 500\,K, 0.05\,dex, and 0.05\,dex, for \teff,
\logg, and \lheh, respectively.

\begin{table}[b]
\begin{center}
\caption[]{Atmospheric parameters of the subdwarf in \target, derived
  from LTE spectral fits}
\label{tbl:specpar}
\begin{tabular}{llll}
\hline \hline \noalign{\smallskip}
Telescope   & \teff      & \logg     & \lheh       \\
            &          K &  cm\,s$^{-2}$ &         \\
\noalign{\smallskip} \hline \noalign{\smallskip}
 NOT        & 27370(90)  & 5.470(13) & --2.654(28) \\
 WHT        & 27990(70)  & 5.528(11) & --2.650(37) \\
\noalign{\smallskip} \hline \noalign{\smallskip}
 adopted    & 27700(300) & 5.50 (3)  & --2.65(3)   \\
\noalign{\smallskip} \hline
\end{tabular} \end{center} \end{table}

Besides the lines of the hydrogen Balmer series and \ion{He}{i} lines,
clear lines of heavier elements are present.  To quantify these, we
fitted the mean WHT spectrum with the NLTE model atmosphere code {\sc
  Tlusty 204} \citep{hubeny95} and performed spectral synthesis with
{\sc Synspec 49} \citep{lanz07}.  Our models included H, He, C, N, O,
Si and Fe opacities consistently in the calculations for atmospheric
structure and synthetic spectra.  We fit the observed spectrum with
the {\sc XTgrid} fitting { code} \citep{nemeth12}.  This procedure is
a standard $\chi^2$-minimization technique, that starts with a
detailed model and by successive approximations along the steepest
gradient of the $\chi^2$ it converges on a solution.  Instead of
individual lines, the procedure fits the entire spectrum so as to
account for line blanketing.  However, the fit is still driven by the
dominant Balmer lines with contributions from the strongest metal
lines (listed in Table~\ref{table:lines}).  We used a resolution of
$\Delta\lambda=1.7$ \AA\ and assumed a non-rotating sdB star.

The best NLTE fit was found with \teff\,=\,28\,100 K and
\logg\,=\,5.55\,dex, using the Stark broadening tables of
\cite{tremblay09}.  When using the line broadening tables from
\cite{lemke97} alone we found a lower surface temperature and gravity,
by about 860 K and 0.07 dex, respectively. A very similar systematic
shift was found for the sdB star \object{KIC\,10553698} by
\citet{ostensen14a}.
 
\begin{table}[b]
\caption[]{Parameters for the NLTE fit shown in
  Fig.\,\ref{fig:meanSpectra}, with solar abundance ratios 
  from \citet{asplund2009} for comparison.}
\label{table:abund}
\centering
\begin{tabular}{lr@{~~~}r@{~~~}rcl} \hline\hline \noalign{\smallskip}
Parameter                            & Value   & $+1 \sigma$ & $-1 \sigma$  & Unit
& Solar\\
\noalign{\smallskip} \hline \noalign{\smallskip}
  \teff\                               & 28100    & 1080  & 220   & K   &           \\ 
  \logg\                               &  5.550   & 0.008 & 0.026 &  cm\,s$^{-2}$ & \\ 
  $\log n(\mathrm{He})/n(\mathrm{H})$\ & --2.57   & 0.03  & 0.07  & dex & --1.07(1) \\ 
  $\log n(\mathrm{C})/n(\mathrm{H}) $\ & --5.00   & 0.40  & 0.40  & dex & --3.57(5) \\ 
  $\log n(\mathrm{N})/n(\mathrm{H}) $\ & --4.14   & 0.09  & 0.13  & dex & --4.17(5) \\ 
  $\log n(\mathrm{O})/n(\mathrm{H}) $\ & --4.59   & 0.24  & 0.30  & dex & --3.31(5) \\ 
  $\log n(\mathrm{Si})/n(\mathrm{H})$\ & --5.79   & 0.34  & 0.51  & dex & --4.49(3) \\ 
  $\log n(\mathrm{Fe})/n(\mathrm{H})$\ & --4.31   & 0.48  & 0.04  & dex & --4.50(4) \\ 
\noalign{\smallskip} \hline
\end{tabular}
\end{table}  

Errors and abundances for the elements that were found to be
significant are listed in Table\,\ref{table:abund}.  Statistical
errors were determined by changing the model in one dimension until
the critical $\chi^2$-value associated with the confidence limit at
the given degree-of-freedom was reached.  The resulting fit is shown
together with the mean spectrum in Fig.\,\ref{fig:meanSpectra}.

Our NLTE model provides consistent parameters with the LTE analysis.
We find that the abundances of Fe and N are about solar, whereas the
other elements are depleted.  This pattern agrees with the typical
abundance profile of sdB stars \citep[see e.g.][]{geier13a}.  High
depletion is normal in sdB stars due to gravitational settling, and
large deviations from the solar mixture for individual elements are
also common \citep{heber00}.

\section{The orbital light curve from \kep\ photometry}

We analysed the \kep\ light curve of \target\ as obtained in
\kep\ quarters Q06--Q17, totalling 2.88 year of data with 58.85\,s
sampling time \citep[ i.e. short-cadence data;
  see][]{gilliland10b}. This photometric time series of 1.4 million
data points spans BJD 2455372 to 2456424, which is roughly the same
range as for our spectroscopic dataset.  As for the similar sdBV+WD
binary \object{KIC\,11558725} \citep{telting12}, we expect that part
of the variability in the \kep\ light curve could be due to orbital
Doppler beaming.

\begin{figure*}[t]
\centerline{\psfig{figure=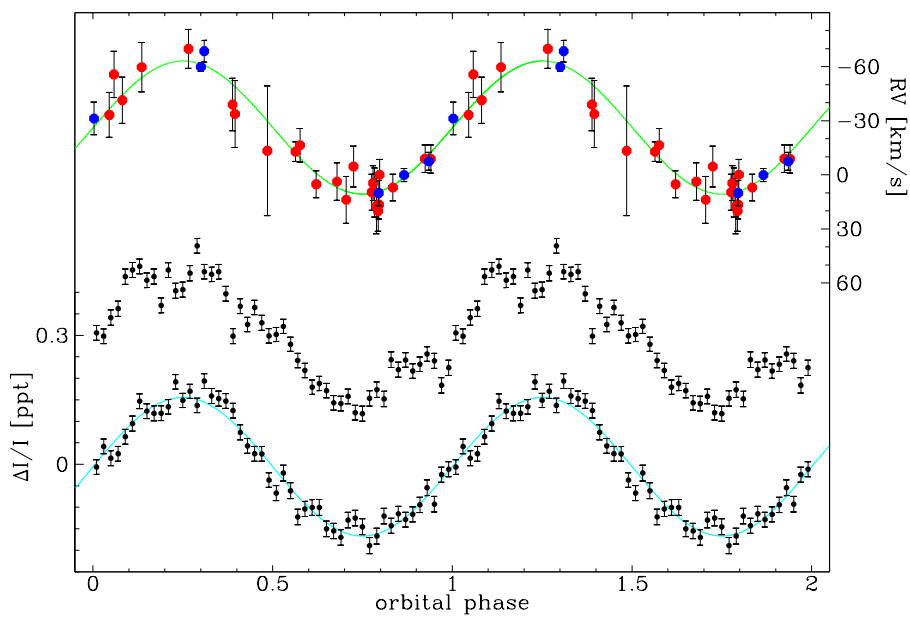,height=12cm,angle=-90}}
\caption[]{Bottom: the first half of the 2.8 year long \kep\ light
  curve from quarters Q06 through Q16, folded on the orbital period and
  binned into 50 bins, with a model fit as listed in Table~\ref{tbl:orbit-lc}.
  Middle:  folded second half of the light curve, offset by 0.3 ppt
  for clarity.  The structure in this curve relates to imperfect detrending
  around gaps in the light curve.
  Top: phased radial velocities from the spectra from the Nordic
  Optical (red dots) and  William Herschel (blue dots) telescopes }
\label{fig:keplerOrbit}
\end{figure*}

\begin{table*}[thb]
\begin{center}
\caption[]{Sine fits to parts of the orbital lightcurve of
  \target\ from \kep\ measurements. $T_{\rm orb,B}$ is half an orbit before
  $T_{\rm orb}$ in Table 1. BJD values have 2450000 subtracted.}
\label{tbl:orbit-lc}
\begin{tabular}{cccccc}
\hline \hline 
\noalign{\smallskip}
 BJD range            & \# data& $A$ [ppt]  & $P$ [d]      & $T_{\rm orb,B}$ [BJD]  & RMS [ppt]\\
\noalign{\smallskip} 
\hline
\noalign{\smallskip} 
 5372.466 -- 5626.812 & 330175  & 0.1658 (56)& 14.081  (14) & 5964.95 (47) & 2.29 \\
 5626.813 -- 5874.790 & 330178  & 0.1632 (55)& 14.164  (15) & 5967.84 (25) & 2.20 \\
 5874.791 -- 6139.383 & 330176  & 0.1514 (55)& 14.234  (16) & 5967.50 (09) & 2.18 \\
 6139.384 -- 6390.969 & 330179  & 0.1453 (56)& 14.122  (16) & 5968.88 (35) & 2.25 \\
\noalign{\smallskip} 
\hline
\noalign{\smallskip} 
 5372.466 -- 5874.790 & 660353  & 0.1625 (39)& 14.1566 (52) & 5967.61 (14) & 2.25 \\
\noalign{\smallskip} 
\hline
\noalign{\smallskip} 
 5372.466 -- 6390.969 & 1320708 & 0.1546 (28)& 14.1666 (19) & 5967.84 (04) & 2.23 \\
\noalign{\smallskip} 
\hline
\end{tabular} \end{center} \end{table*}

The 2.88 year coverage has 37 gaps in the data, with durations
typically on the order of one day due to safing and repointing events,
but including five prolonged periods ($\sim$5 to $\sim$15 days) void
of data.  The data typically needs detrending to account for
instrumental offsets and drifts following data voids.  Given that the
binary period of \target\ is close to that of the spacing of the data
gaps, we attempted to process the raw \kep\ data in such a fashion to
avoid any unnecessary loss of amplitude at the binary period.

First, we checked for possible contamination by neighbouring
stars. Fortunately, the signal in the optimal aperture chosen by the
standard pipeline is only slightly affected by that of other objects
around our target (see the seasonal contamination described in the
next subsection).  This made the data processing less complex as we
could avoid customization of the optimal aperture. Subsequently, we
tried out fluxes from the pre-search data-conditioning (PDC) and
simple aperture-photometry (SAP) apertures \citep{jenkins10}. The PDC
fluxes were nicely detrended, however the data stitching implicitly
removed the binary signal. As we could not use the PDC fluxes, we
worked with SAP fluxes only, which however still needed to be
corrected with some kind of detrending. For this purpose we used
cotrending basis vectors (CBV). The PyKE \citep{kinemuchi12} pipeline
allows the implementation of a number of CBVs to approximate the
systematics seen in the light curve. Often, we have to apply
additional detrending by means of spline fitting to the residuals.  In
order not to affect the amplitude of the binary signal we skipped that
step in this particular case, allowing some long-term instrumental
variation to be present in our data. Finally, the data were
4.5$\sigma$ clipped to remove outliers and were transformed to
fractional intensities $\Delta$I/I.  The full light curve of 1.4
million points has a standard deviation $\sigma$=2245ppm\,.  For the
determination of the photometric orbital amplitude we only used the
1.32 million zero-flagged \kep\ data points of Q06--Q16, with standard
deviation $\sigma$=2235ppm\,.

We find that the applied trend correction works almost perfectly in
the first half of the data set, but leaves obvious offsets around many
data voids in the latter half of the data set.  This may also be
aggravated by possible deterioration of the CCDs over the lifetime of
\kep, and the lack of updated flatfields to account for this.  In
Fig.~\ref{fig:keplerOrbit} we show the \kep\ light curve folded on the
orbital period into 50 phase bins, for both the first and second half
of the data, showing that the latter folded curve shows significant
deviations induced by imperfect detrending.

The level of detail to which this affects the determination of the
photometric orbital amplitude is presented in Table 4, where we list
the results of sine fits to parts of the processed \kep\ data of
\target\ .  Here we divided the Q06--Q16 data set in 4 parts of equal
number of data points.  Whereas the amplitude at the orbital period is
consistent for the first two quarters, it seems to decline in the last
two quarters.

For this reason we use only the first half of the \kep\ data to
determine the photometric orbital parameters of \target.  A sine fit
reveals the orbital period to be 14.1566(52)\,d, in agreement with the
period found from spectroscopy, with amplitude of 162.5(3.9)\,ppm.  We
do not find orbital harmonics.

\subsection{Doppler beaming}

The high precision \kep\ data permits us to accurately explore the
low-amplitude Doppler beaming effect, something that is very hard to
do with ground-based data. This effect is induced by stellar motion in
a binary orbit and causes brightness modulation \citep{RL79}.  The
Doppler beaming effect permits an estimate of the radial velocity
without resorting to spectroscopic data.  A confirmation of the
correspondence between Doppler beaming amplitudes, radial velocities,
and the light-time effect, all in one object, was presented by
\cite{telting12}.

The light curve of \target\ displays a brightening of the sdB star at
the orbital phases where the star is approaching us in its orbit
(i.e.~when its orbital radial velocities are negative, see
Fig.~\ref{fig:keplerOrbit}), and this is exactly the effect expected
by Doppler beaming. Note that unlike \object{KPD\,1946+4330}
\citep{bloemen11}, \target\ does not show any sign of ellipsoidal
deformation which in the closest sdB+WD binaries produces a strong
signal at \porb/2 \citep[see also e.g.][]{silvotti12}.  This is
consistent with the much longer orbital period of \target\ as opposed
to the 0.4\,d period of \object{KPD\,1946+4330}.

In the case of Doppler beaming the observed flux from the target,
$F_\lambda$, is related to the emitted spectrum, $F_{0,\lambda}$, and
the projected orbital velocity, $v_r, as$
\begin{equation}
F_\lambda = F_{0,\lambda} \left( 1 - B {v_r\over c} \right)
\end{equation}
and the beaming amplitude relates to the orbital radial-velocity
{ amplitude, $K$, as}
\begin{equation}
A_B ~ D = B {K\over c} ~ ,
\end{equation}
where $ A_B $ is the amplitude of the beaming signal in the light
curve (see Fig. 3), and $D$ the \kep\ decontamination factor.  The
beaming factor $B$ is reflecting the relativistic aberration of the
wavefront and the Doppler shift of the target spectrum within the
\kep\ wave band. It was computed for \object{KIC\,11558725} by
\cite{telting12}, following the procedure described in
\citet{bloemen11}.  Here we adopt the same value $B$=\,1.403(20),
since \object{KIC\,11558725} and \target\ have very similar effective
temperature and surface gravity.

The \kep\ pixels onto which our target is imaged suffer from
contamination from neighbouring objects, and from passing charge from
brighter sources when clocking out the CCD.  According to the
\kep\ Archive Handbook, the seasonal contamination factors for
\target\ are 0.048, 0.067, 0.066, 0.041 or 0.0555 on average, implying
that all periodic amplitudes derived from these \kep\ fluxes should be
multiplied by $D$\,=\,1/(1$-$0.0555)\,=\,1.059 to get the intrinsic
amplitudes of \target.

We find that the photometric amplitude of the Doppler beaming as seen
by \kep\ and corrected for contamination, is consistent with the
spectroscopically derived radial-velocity amplitude within the errors
of the data.  Given the observed amplitude $A_B$=163(4)\,ppm (Table
4), we use Eq.\ (2) to derive a value of $K$=36.8(1.1)\,km\,s$^{-1}$
for the orbital radial-velocity amplitude, from Doppler beaming.

The fact that this value is consistent with that of the spectroscopic
value proves that the companion of the sdB does not significantly
contribute to the observed Doppler beaming, consistent with a compact
nature of the companion. Similarly, as the beaming fully accounts for
the photometric variability at $P_{\rm orb}$, and no reflection-effect
variability is observed, the companion must be of a compact nature.

\section{The orbit derived from the light travel-time probed by the pulsations}

In analogy to the effect that the Earth's orbit has on arrival times
of a variational signal, and on the apparent frequencies of that
signal, the orbit of the sdB alters the phases and
frequencies of its pulsations as perceived by a distant observer.

For the similar sdB+WD system \object{KIC\,11558725} \citep{telting12}
we have shown that it is possible to derive an independent orbital
solution by using the pulsations as clocks to exploit the orbital
light travel-time delays, equivalent to the method used by
\citet{hulse75} for pulsars.  If the orbit is not known, one can use
the pulsations as clocks to derive the light-travel time that
corresponds to the radius of the orbit, which relates to the
radial-velocity amplitude $K$ that can be obtained from spectroscopy
or the Doppler-beaming curve as
\begin{eqnarray}
\Delta t_{\rm R} = \frac{a_{\rm sdB} \sin i }{c}= \frac{K}{c} \frac{P_{\rm orb}}{2\pi}
\sqrt{1-e^2} ~ ,
\end{eqnarray}
where we introduce the R\o mer delay, $\Delta t_{\rm R}$, to represent
the light-travel time.
For a circular orbit, the light-travel delay as a function of the
subdwarf's position in its orbit can be written as
\begin{eqnarray}
T_{\rm delay}(t) =  \Delta t_{\rm R} ~
\cos\left(\frac{2\pi}{P_{\rm orb}} (t - T_{\rm orb,R})\right) ~ ,
\end{eqnarray}
where $T_{\rm orb,R}$ is the time at which the subdwarf is closest to
the Sun in its orbit, corresponding to that listed in Table 1.

We did not detect any second- or third-harmonic peaks, neither for the
orbit nor for the pulsations.  Hence, for sinusoidal signals the
\kep\ light curve of \target\ can be approximated by
\begin{eqnarray}
\frac{\Delta I(t)}{I} & = &  A_B  \sin \left( \frac{2\pi}{P_{\rm orb}}(t - T_{\rm
  orb,B}) \right)  \nonumber \\
 & + &  \sum_{i} A_{\rm i,puls}  \sin \left( 2\pi F_{\rm i,puls} 
(t - T_{\rm i,puls} + T_{\rm delay}(t) ) \right)   ~ ,
\end{eqnarray}
where the first term describes the orbital beaming effect, and where
all individual pulsations are affected by the same orbital
light-travel delay $T_{\rm delay}(t)$.  Here, the phase of the
individual pulsations, $T_{\rm i,puls}$, is expressed in the time
domain rather than as an angle.  Note that the above sum of sine
curves is equivalent to the model that we fit as part of the
prewhitening procedure described in the next section, with the
addition of a phase delay that introduces just two extra parameters,
i.e. the amplitude $\Delta t_{\rm R}$ and timing reference 
$T_{\rm orb,R}$ of the light-travel delay.

We apply the above Eq.\ (5) to derive a third independent measurement
of the radial-velocity amplitude of the subdwarf in \target.  For this
purpose we fitted the above model to the \kep\ Q06--Q17 light
curve. For the light-travel delay and corresponding timing zero point
we use separate parameters from those representing the
Doppler-boosting amplitude and timing zero point in the first term of
Eq.\ (5), in order to make this determination of the orbital
parameters truly independent from the Doppler-boosting value.  In the
model we included the 70 strongest frequencies of the subdwarf, with
amplitude threshold of S/N\,$>$\,6.4.  We repeated this using only the
strongest 30 (S/N\,$>$\,9.4), and only the strongest 10 pulsations
(S/N\,$>$\,38).

\begin{table}[h]
\begin{center}
\caption[]{Orbital light travel-time delay in 
           \target\ from timing the pulsations}
\label{tbl:orbit-ltt}
\begin{tabular}{ccc}
\# of pulsations & $\Delta t_{\rm R}$ &  $T_{\rm orb,R}$ \\
\hline \hline \noalign{\smallskip}
10 &   23.8  (3.8)&    5975.31 (0.37) \\
30 &   25.6  (2.7)&    5974.89 (0.24) \\
70 &   26.5  (2.5)&    5974.95 (0.21) \\
\noalign{\smallskip} \hline
\end{tabular} \end{center} \end{table}

\begin{figure*}[t]
\centerline{\psfig{figure=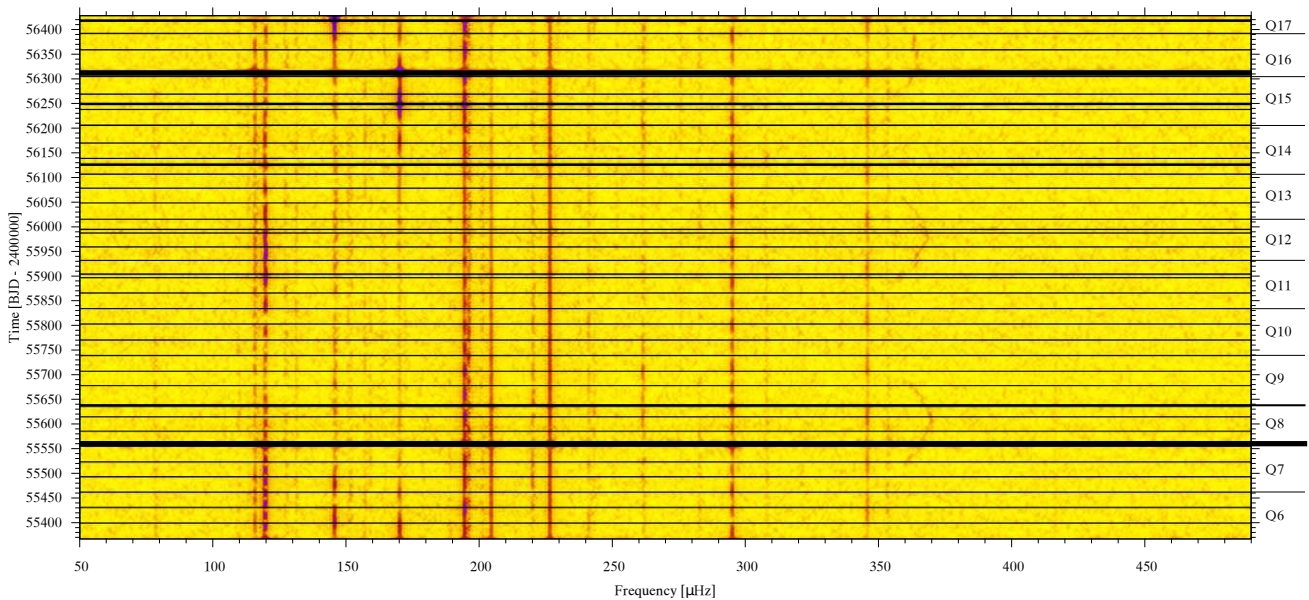,height=8cm}}
\caption[]{Dynamic Fourier spectrum from quarters Q06 through Q17
  computed for 10 day stretches of data, versus BJD -- 2400000.
  The frequency-variable feature at 365 $\mu$Hz is a spacecraft
  artefact \citep{baran13}.
  The dark horizontal lines reflect data gaps.
} 
\label{fig:runningFT}
\end{figure*}

Simultaneously fitting all 70 pulsational amplitudes, frequencies and
phases, together with the R\o mer timing zero point and delay as free
parameters, we find $T_{\rm orb,R}$=5974.95(21)\,[BJD -- 2450000] and
$\Delta t_{\rm R}$=26.5(2.5)\,s. The latter is equivalent to a value
of $K$=40.8(3.8)\,km\,s$^{-1}$ for the orbital radial-velocity
amplitude. Both the R\o mer timing zero point and delay are in
excellent agreement with the spectroscopic values in
Table~\ref{tbl:orbit}, and with those derived from the Doppler
beaming.  We note that the accuracy of the light-travel delay can be
improved by using many pulsations in the model, even if they have
amplitudes with only moderate S/N, as can be judged from
Table~\ref{tbl:orbit-ltt}.

\section{The pulsation spectrum}

After having used the pulsations as clocks to confirm the orbit, we
here present the pulsation frequencies in \target\ in detail.  We
thereafter proceed to present mode identifications that will be useful
input for a possible seismic study of \target. In this paper we
specifically aim to derive the rotation period of the star as a first
seismic result.

For the determination of the pulsational signal in the \kep\ light
curve of \target\ we used the full Q06--17 data set described in
Sect.\ 3, of in total 1.4 million points.  The orbital light-time
effect generates orbital sidelobes in the pulsational frequency
spectrum \citep[see e.g.][]{SK12,telting12}; for
\target\ the amplitude of the orbital sidelobes are at the level of
the noise in Fourier space.  We opted to first transform the
observation timings to remove the orbit, equivalent to the removal of
Earth's motion in the heliocentric timing correction.  While this
transformation removes the orbital sidelobes from the frequency peaks,
it will also focus all power of the sidelobes into that of the main
frequency peak, slightly enhancing the S/N of the intrinsic
frequencies.

An iterative prewhitening process, involving an implementation of the
fast-Fourier transform that allows for unevenly sampled data
\citep[see][page 574]{press1992} to find peaks, and subsequent
non-linear least-squares (NLLS) sine-curve fits to subtract the
pulsational content mode by mode from the original data, was used to
derive the frequency list of Table~\ref{tbl:freqlist}.  We extracted
132 significant frequencies, in addition to the orbital frequency.

We use the mean value of the Fourier amplitude spectrum of the
original data to approximate the standard deviation on the amplitudes
in Fourier space, $\sigma_{\rm FT}$=3.2\,ppm.  The mean value of the
Fourier amplitude spectrum after removing the orbital and pulsational
variations is still 3.2\,ppm, which shows that the variance of the
original data is dominated by measurement errors, and not by signal
from our target. We adopt 4.5$\sigma_{\rm FT}$=14.4\,ppm as the
threshold of significance of the peaks in the Fourier amplitude
spectrum.  At the level of this 4.5$\sigma$ threshold we expect about
5 spurious peaks below the Nyquist frequency of 8500\,$\mu$Hz, given
the 1/$T$ resolution of 0.011\,$\mu$Hz.  Table~\ref{tbl:freqlist}
lists the 132 significant pulsation frequencies, with 16 other
lower-significance frequencies that seem convincing-enough multiplet
members to be listed (3.4 $<$ S/N $<$ 4.5).

The errors on the pulsational parameters as listed in
Table~\ref{tbl:freqlist} were obtained by fitting a non-linear
multi-sine model, while using the standard deviation
$\sigma$=2245\,ppm of the \kep\ data as an estimate of the measurement
errors. This results in a reduced $\chi^2$ value of 0.94, with an RMS
of 2180\,ppm, and an estimate of the formal errors on the pulsation
amplitudes of 2.6\,ppm.  The latter is somewhat smaller than our error
estimate of 3.2\,ppm based on the mean (residual) Fourier amplitude
spectrum, and we have used the more conservative of the two error
estimates to test the significance of individual peaks, and for the
S/N column in Table~\ref{tbl:freqlist}.  We stress that for many of
the modes the amplitude variability is larger than the formal errors
on their fitted amplitude.

We note that the observed amplitudes in Table~\ref{tbl:freqlist} have
neither been corrected for the \kep\ decontamination factor $D$=1.059
(see Sect.\ 3), nor for amplitude smearing due to the effective
exposure time of 58.85 seconds.  Both effects cause the observed
amplitudes to be smaller than the intrinsic amplitudes. The latter
affects mostly the shortest periods ($p$-modes): for the shortest
detected periods of 211 seconds the observed amplitude is reduced by
12\%.

It is clear from Table~\ref{tbl:freqlist} that most pulsation
frequencies lie in the $g$-mode domain, while only few low-amplitude
\mbox{$p$ modes} are present in \target .  Most pulsations are not
stable over time, which causes prewhitening to be difficult, and which
otherwise manifests itself as a bunch of narrow peaks around most of
the pulsation peaks in the Fourier spectrum (see Figs.\ 5, 6, 9).  To
investigate for each pulsation the magnitude of any amplitude or
frequency variations, we list the amplitudes of the pulsations for the
first and second half of the data set, fitted while keeping fixed
frequencies. The resulting amplitudes are listed in
Table~\ref{tbl:freqlist}.

To illustrate the fact that most strong pulsation modes are present
throughout the \kep\ run, we show a section of the Fourier transform
in a dynamic form in Fig.\ 4.  The multiplets show clear beating
patterns in this figure.  It appears that some modes are relatively
stable throughout the 3 years of \kep\ monitoring, while other modes
show significant amplitude variability on timescales comparable to or
even longer than the duration of the Kepler run. This is in contrast
with the much shorter $\sim$60\,day timescale of the amplitude
variability of the stochastic pulsations seen in \object{KIC\,2991276}
\citep{ostensen14b}.

On the other hand there are many modes in \target\ that show amplitude
variability with prolonged periods where the mode is not excited.
Some modes are of short-lived nature, with life times comparable to
the rotation period (see Sect.~6); some permanently present modes have
short-lived episodes of high amplitude.



\section{Pulsational period spacings and multiplet splittings,
  rotation and inclination}

Recently, \citet{reed11c} have revealed that $g$-modes in sdB stars
show sequences with near-constant period spacings.  As the period
spacings follow the asymptotic relation for $g$-modes, the spacings
relate to the value of the spherical-harmonic degree $\ell$ of the
modes, and hence allow us to determine the $\ell$-value of the modes
directly.  \citet{reed11c} already identified $\ell$ values for the
$\sim$20 pulsations that could be resolved in the \kep\ survey data of
\target.  They identified an $\ell$=1 sequence with period spacing
$\Delta \Pi_{\ell=1}$=248\,s, and an $\ell$=2 sequence with period
spacing $\Delta \Pi_{\ell=2}$=145\,s, with a ratio close to the
theoretically expected $\sqrt{3}$.  For the modes identified through
the period sequence the radial order $n$ can be determined directly,
save for a zero-point offset that has to be modelled.  For our
identification of the radial order $n$ of the modes in
Table~\ref{tbl:freqlist} we set the zero-point to modes with periods
that satisfy $P\sqrt{\ell(\ell+1))}\sim$800\,s
\citep[see][]{charpinet02}.

To the first order in the rotation frequency $\Omega$, the observed
frequencies of the modes are altered by $m \Omega (1 - C_{n\,\ell})$,
with $m$ the azimuthal quantum number of the spherical harmonic.
Hence, for non-radial modes of given degree $\ell$, we expect
frequency multiplets of 2$\ell$+1 peaks to occur.  In the asymptotic
regime, the value of $C_{n\,\ell}$ can be approximated by 0 for
$p$-modes, and by $1/(\ell(\ell+1))$ for high-$n$ $g$-modes.  The
latter implies a value of $C_{n\,\ell}$$\sim$0.5 and
$C_{n\,\ell}$$\la$0.17 for $\ell$=1 and $\ell$>1 $g$-modes,
respectively.

The (multiplet) frequency splittings we discuss below are derived
by taking the frequency difference of subsequent entries in the sorted
frequency list in Table~\ref{tbl:freqlist}.

\renewcommand{\textfraction}{0}
\renewcommand{\dbltopfraction}{1}

\begin{figure}[t]
\centering
\psfig{figure=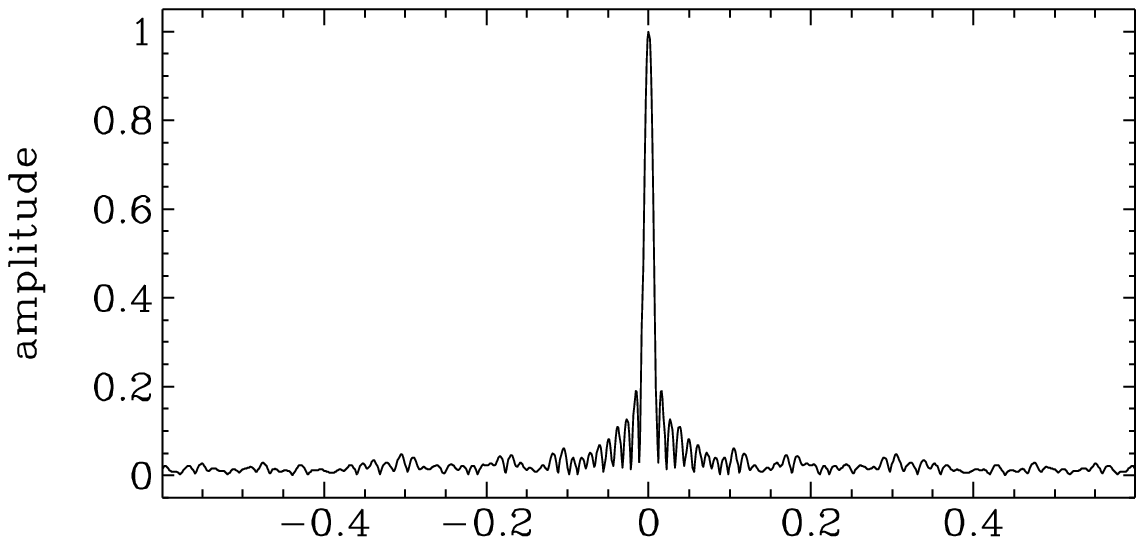,width=8.20cm}
\\[2mm]
\psfig{figure=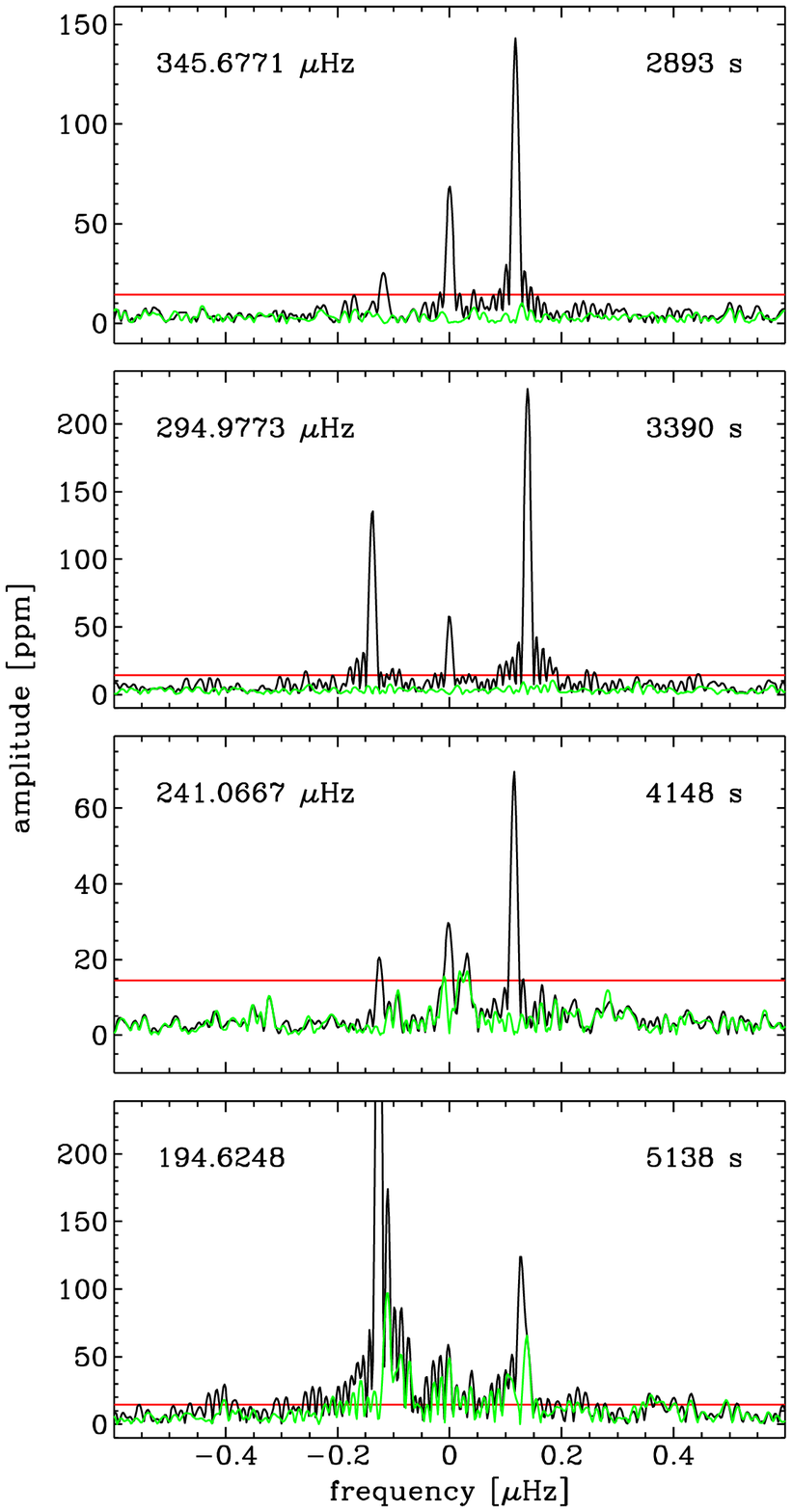,width=8.20cm}
\caption[]{
  { Top: DFT window function of the Q06--Q17 data set.
  ~~~ Bottom panels:}
  Complete and resolved $\ell$=1 triplets in the Fourier
  Transform of the Q06--Q17 data set. The prewhitened FT is overdrawn
  (green); for the central mode in the lowest panel no prewhitening
  was attempted prompted by the amplitude-variable nature of this
  triplet.  The red horizontal line marks the 4.5\,$\sigma$ limit. The
  triplet spacings are 0.12--0.14\,$\mu$Hz.
}
\label{fig:multiplets}
\end{figure}

\begin{figure}[h]
\centering
\psfig{figure=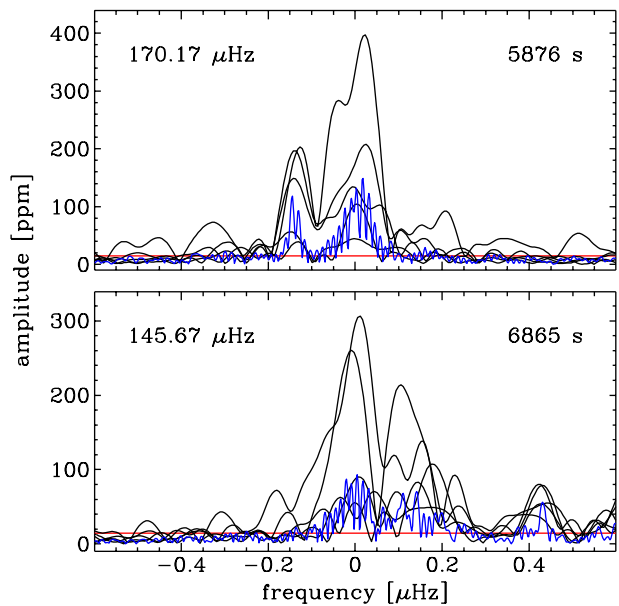,width=8.2cm}
\caption[]{Wide-peaked $\ell$=1 doublets in the Fourier
  Transform of the Q06--Q17 data set (blue). The black curves
  are FTs of the five $\sim$210-day subsets of the data, which show 
  the amplitude variability of these modes.
  The red horizontal line marks the 4.5\,$\sigma$ limit. 
  The doublet spacings are consistent with those in Fig.\ 5.
}
\label{fig:doublets}
\end{figure}

\begin{figure}[h]
\centering
\psfig{figure=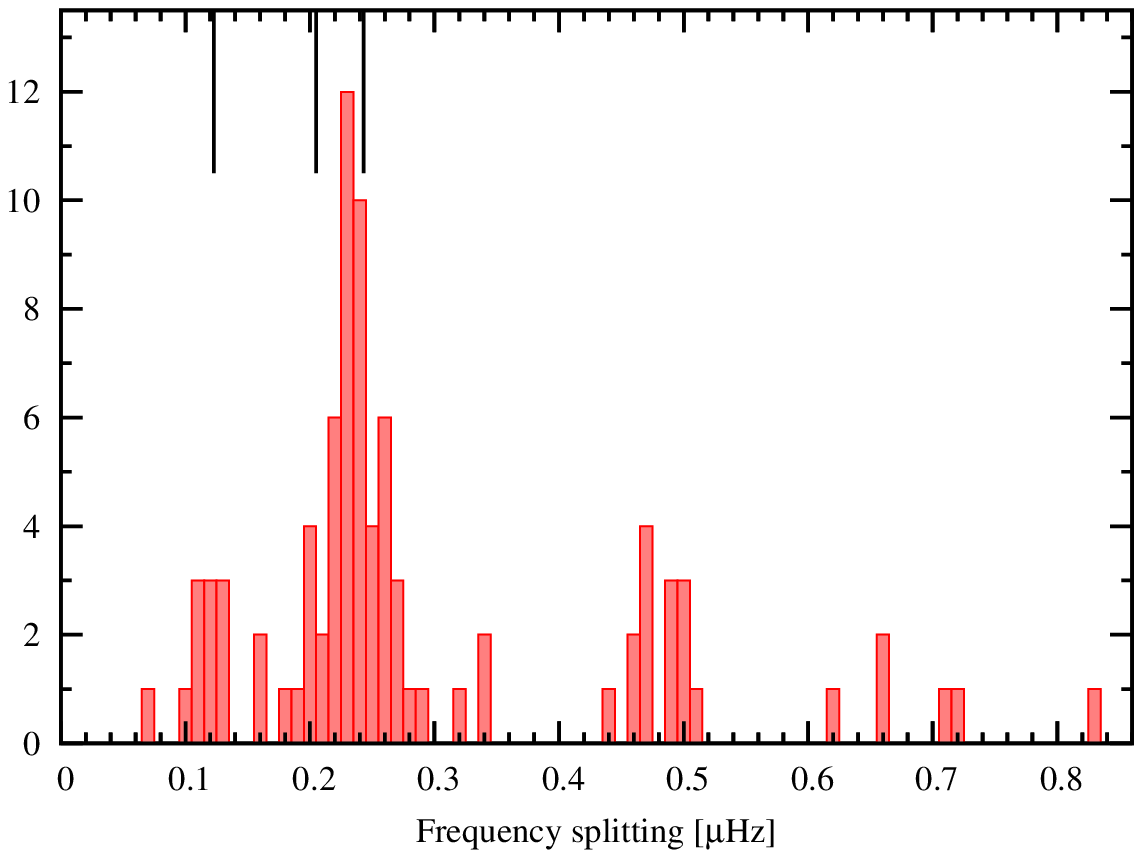,width=9cm}

\caption[]{Histogram of frequency splittings listed in
  Table~\ref{tbl:freqlist}, that are determined by taking the
  frequency difference between subsequent entries in the sorted
  frequency list.  The vertical indicators at 0.123\,$\mu$Hz,
  0.205\,$\mu$Hz, and 0.243\,$\mu$Hz, reflect expected $\ell$=1, 2, 8
  splittings for a rotation period of 47 days.  The splittings at
  0.13\,$\mu$Hz are identified as $\ell$=1, and the peak at
  0.23\,$\mu$Hz as $\ell$$\ge$2 modes.  }
\label{fig:splithisto}
\end{figure}

\subsection{Mode identification: $\ell$=1}

As a base-line observation for mode identification in \target\ we use
the set of four $g$-mode triplets with narrow frequency splittings of
0.12--0.14\,$\mu$Hz, displayed in Fig.~5.  These are the only 
triplets with such narrow splitting in \target. All four triplets are
consistent with the $\ell$=1 period spacing sequence, and all have at
least one frequency peak that is among the strongest peaks in \target.

\begin{figure}[t]
\centering
\psfig{figure=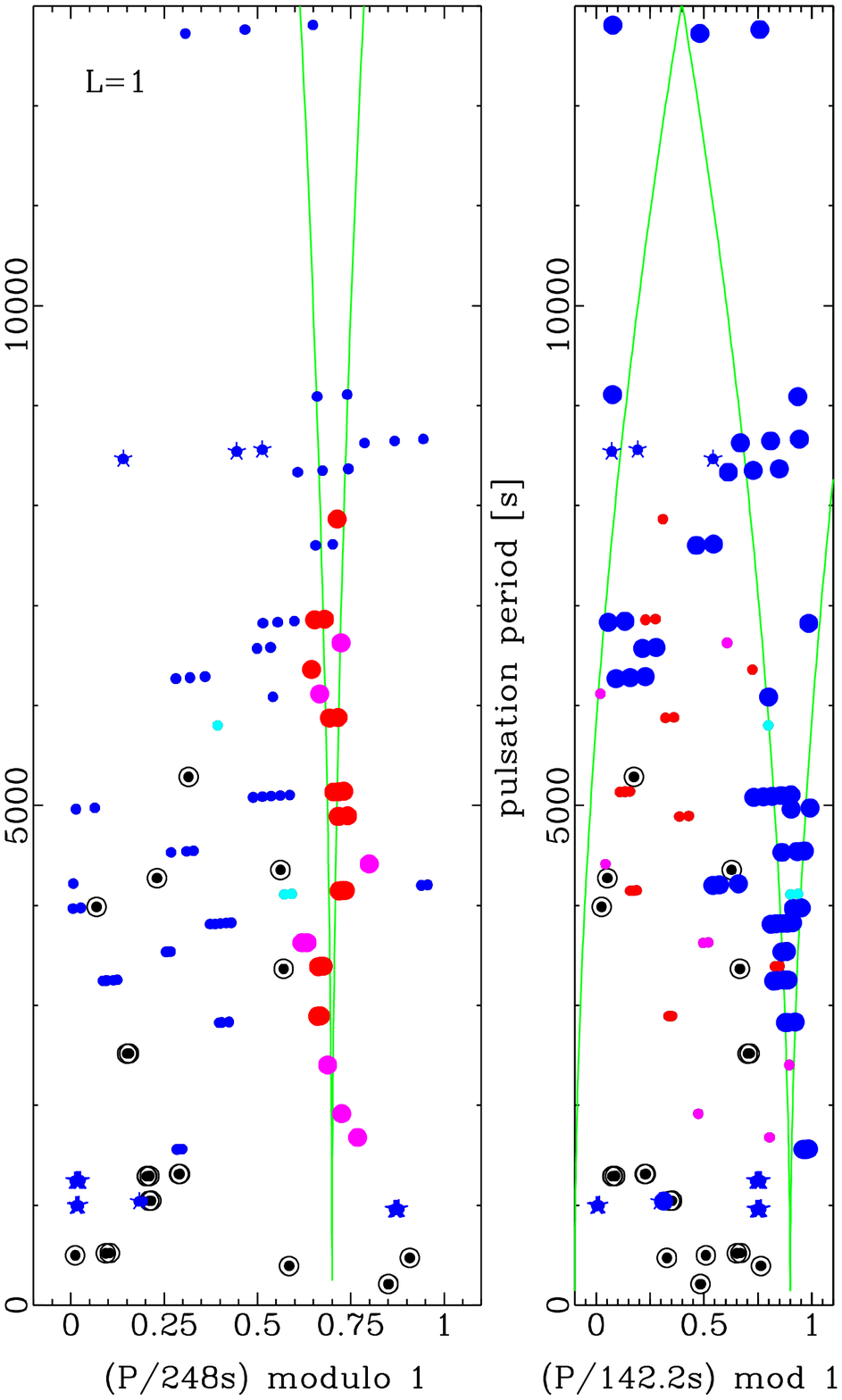,width=9cm}
\caption[]{ Echelle diagrams of the $\ell$\,=\,1 (left) and
  $\ell$\,=\,2 (right) period-spacing sequences.  Green curves reflect
  the full width of $\ell$\,=\,1 (left) and $\ell$\,=\,2 (right)
  multiplets.  Red dots are secure $\ell$=1 identifications from
  frequency splittings; magenta dots are likely $\ell$=1
  identifications based on period spacing.  All dark blue symbols
  reflect modes that have frequency splittings corresponding to
  $\ell$$\ge$2.  Blue stars correspond to modes in $\ell$$\ge$3
  multiplets.  Cyan dots are likely $\ell$=2 identifications based on
  period spacings only.  Ringed dots are modes that could not be
  classified with any certainty.
}
\label{fig:echelle}
\end{figure}

Accepting these triplets as $\ell$=1 modes, the frequency splitting
immediately discloses a rotation period between 41 and 48 days,
assuming $C_{n\,\ell}$=0.5\,.  The observed $\ell$=1 triplet frequency
splitting also implies that multiplet splittings for modes with higher
$\ell$ values can range between 0.20--0.28\,$\mu$Hz assuming
$C_{n\,\ell}$$\la$0.17\,. In Fig.\ 7 we show a histogram of the
multiplet splittings listed in Table~\ref{tbl:freqlist}; the histogram
clearly shows an abundant presence of higher-$\ell$ splittings peaking
at 0.23\,$\mu$Hz, which is consistent with those implied by the
observed $\ell$=1 splitting.

\begin{figure*}[t]
\centering
\psfig{figure=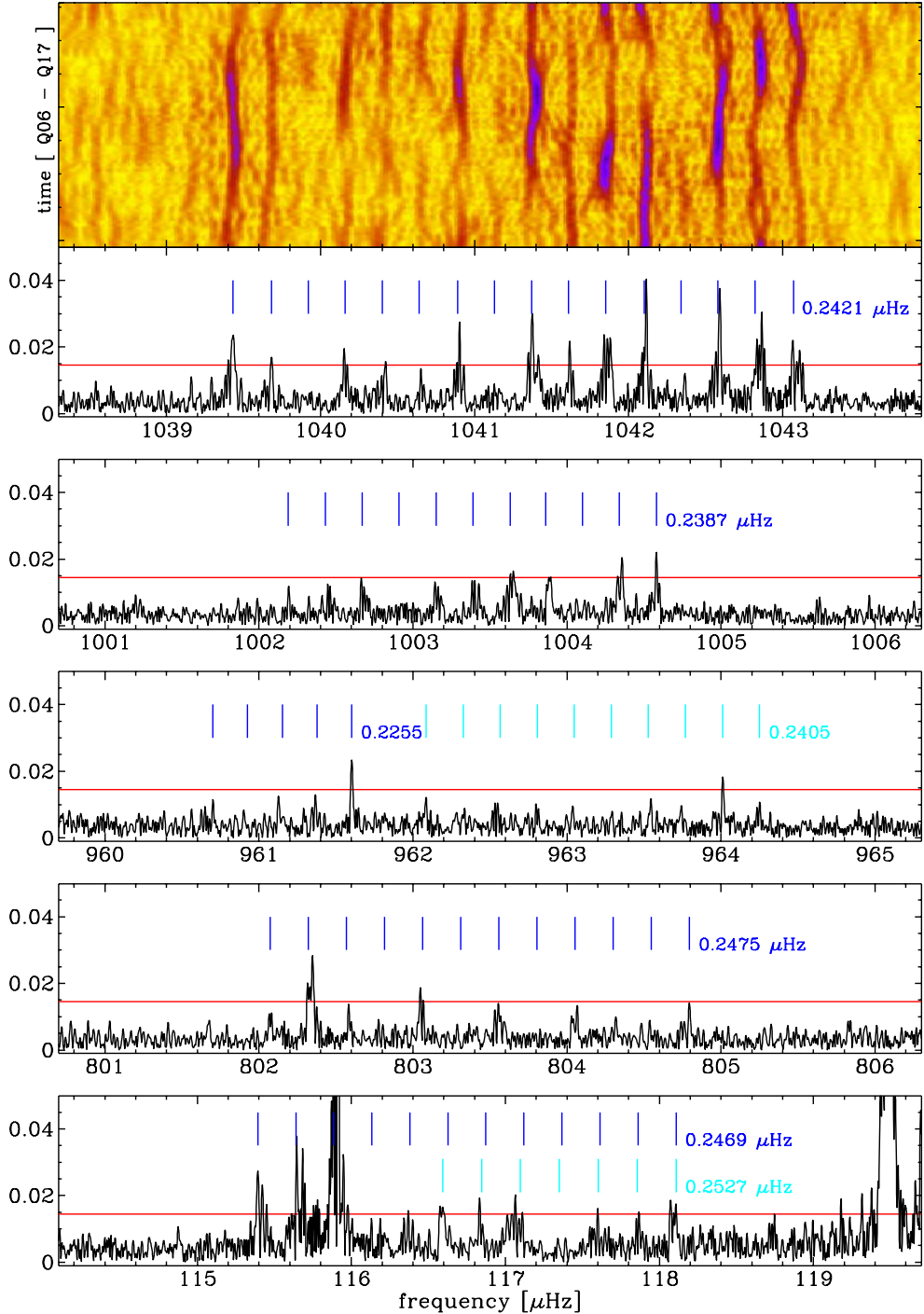,width=13cm}
\caption[]{ Selected regions in Fourier space where large multiplets
  occur, indicative of $\ell$$>$2.  The Fourier amplitudes are given
  in units of ppt; the red line reflects our adopted
  detection limit.  For each apparent sequence of
  modes, a grid of vertical lines is drawn with an indicated average
  splitting based on the frequencies of the first and last clear mode
  in the sequence.  The bottom panel shows either a chance alignment
  of an $\ell$=2 and an $\ell$=3 multiplet, or a single multiplet of
  $\ell$$\ge$6.  The 802--805\,$\mu$Hz structure could be an
  $\ell$$\ge$6 multiplet.  The 1002--1004\,$\mu$Hz structure could be
  an $\ell$$\ge$5 multiplet.  For the 1039--1043\,$\mu$Hz structure,
  which is consistent with an $\ell$$\ge$8 multiplet, the dynamic Fourier
  transform as a function of time is shown; for this a sliding window
  of 200 days was used.    Note that all frequency splittings of these
  $\ell$$>$2 modes are very similar in value.  }
\label{fig:complex}
\end{figure*}

In Table~\ref{tbl:freqlist} we list more $\ell$=1 identifications,
some among the highest-amplitude peaks in the Fourier spectrum, that
were identified from the period-spacing sequence anchored by the 4
triplets in Fig. 5.  Following the period-spacing sequence, we
discovered in the FT a number of likely $\ell$=1 modes that have broad
peaks that reflect short-lived episodes of very high amplitude.  An
example of such a mode is the 145.71\,$\mu$Hz mode labelled f$_5$ in
\cite{baran11b}, which was the strongest mode in the initial 30 days
of \kep\ survey data, but is only seen at high amplitude in the very
beginning and very end of the Q06-Q17 observations.  In Fig.\ 6 we
present examples of these extreme amplitude-variable $\ell$=1 modes,
where the full data set has been divided into 5 near-equal chunks to
bring out the Fourier amplitude variability.  These wide $\ell$=1
peaks are found at low frequencies: strong ones at 145.67 and 170.17
\,$\mu$Hz (see Fig.\ 6), and weak but detectable ones at 127.14 and
157.23 \,$\mu$Hz.

The clear $\ell$=1 period spacing sequence is presented in echelle
format in Fig.~8, where we used the spacing $\Delta
\Pi_{\ell=1}$=248\,s to fold the periods.  Unlike for the sdB+WD
binary \object{KIC\,10553698} \citep{ostensen14a}, the pulsations in
\target\ are not excited for many subsequent values of the radial
order $n$.  Although for radial orders $\sim$4--25 (periods
1650--6900\,s) the sequence is followed quite well, the longest
continuous string of excited $\ell$=1 modes is only 5 orders long:
orders $\sim$21--25 (see Table~\ref{tbl:freqlist}).  Hence it is not
feasible to conclusively identify trapped modes
\citep{charpinet02,ostensen14a}, from the $\ell$=1 and $\ell$=2 period
sequences in \target.  Note however, that there {\em are} modes that
do not fall on either of these period sequences (see Fig.\ 8), and
these may be either trapped $\ell$=1 or $\ell$=2 modes or modes with
even higher degree $\ell$.  The only high-amplitude modes among these
are all shown in Fig.\ 9 and may all be part of high-degree
multiplets.

\subsection{Mode identification: $\ell$=2}

As will be shown below, there is strong evidence for a few multiplets
of more than 5 components in \target.  As the frequency splitting for
all $\ell$$\ge$2 degree multiplets is expected to be similar, and
because $\ell$$>$2 multiplets are observed, it is inconclusive to
label a quintuplet with an $\ell$=2 origin, as it could be an
incomplete multiplet of higher degree.

In the right-hand panel of Fig.~\ref{fig:echelle} we show the
echelle diagram for a folding period of 142.2\,s, which is close to
the theoreticaly expected $\Pi_{\ell=1}/\sqrt{3}$=143.2\,s period
spacing. In Table~\ref{tbl:freqlist}, we list the most likely $\ell$=2
modes, based on frequency splitting and period spacing.  A group of 9
multiplets clearly matching the period spacing is found for periods
between 2800--5100\,s (radial orders 17--33), although the longest
continuous string of excited $\ell$=2 modes is only 3 orders long.
Note that in Figure 8 we denote the $\ell$$\ge$3 multiplets with star
symbols.

\subsection{Curiosity cabinet: modes with degree $\ell$$\ge$3}

Even though high-degree modes have low visibility due to cancellation
effects on the stellar disk, there is strong evidence that such modes
are visible in \target\ to the sensitive eye of the \kep\ spacecraft.
\cite{dziem77} and \cite{SW81} list the cancellation factors as a
function of $\ell$ for certain choices of the limb-darkening
law. Photometric observability of modes with degrees
4$\le$$\ell$$\le$8 is expected to drop by two orders of magnitude with
respect to that of $\ell$=1 modes.  In case the photometric
variability has a significant component due to surface-gravity
effects, the drop may be only of one order of magnitude.  As the
highest $\ell$=1 amplitudes in \target\ are around 400 ppm,
intrinsically strong high-$\ell$ modes may just be detectable by \kep.
 
In Figure 9 we show 5 frequency ranges in which we find evidence for
3$\le$$\ell$$\le$8 multiplets in \target. None of these multiplets is
fully complete, but in some only few components are missing.  All
these high-$\ell$ multiplets show the frequency splitting of
0.239--0.253\,$\mu$Hz.  At 1002--1005\,$\mu$Hz we find an $\ell$=6
multiplet, with most peaks barely at the detection limit, but which
is nonetheless a convincing multiplet due to the distinct pattern of
the peaks. At 1039--1043\,$\mu$Hz we find an $\ell$=8 multiplet, with
11 clearly detected modes, and an average splitting of 0.242\,$\mu$Hz.
As for these high-$\ell$ modes the value of $C_{n\,\ell}$ is expected
to be close to zero, the rotation period we derive from the
splittings ranges between 45.7--48.4\,d.

We also show a dynamic Fourier transform of the 1039--1043\,$\mu$Hz
region, based on 200-day chunks of \kep\ data. This diagram shows that
both amplitudes and frequencies are not stable over time. The
individual splittings between subsequent components change over time,
and only through the extensive length of the data set can accurate
average splittings be determined.  Like some of the $\ell$=1 modes,
some modes in these high-$\ell$ multiplets are relatively short lived.
It seems that to find the missing components of the multiplets
\kep\ should have performed on the same field for only a bit longer.

\begin{figure}[t]
\centerline{\psfig{figure=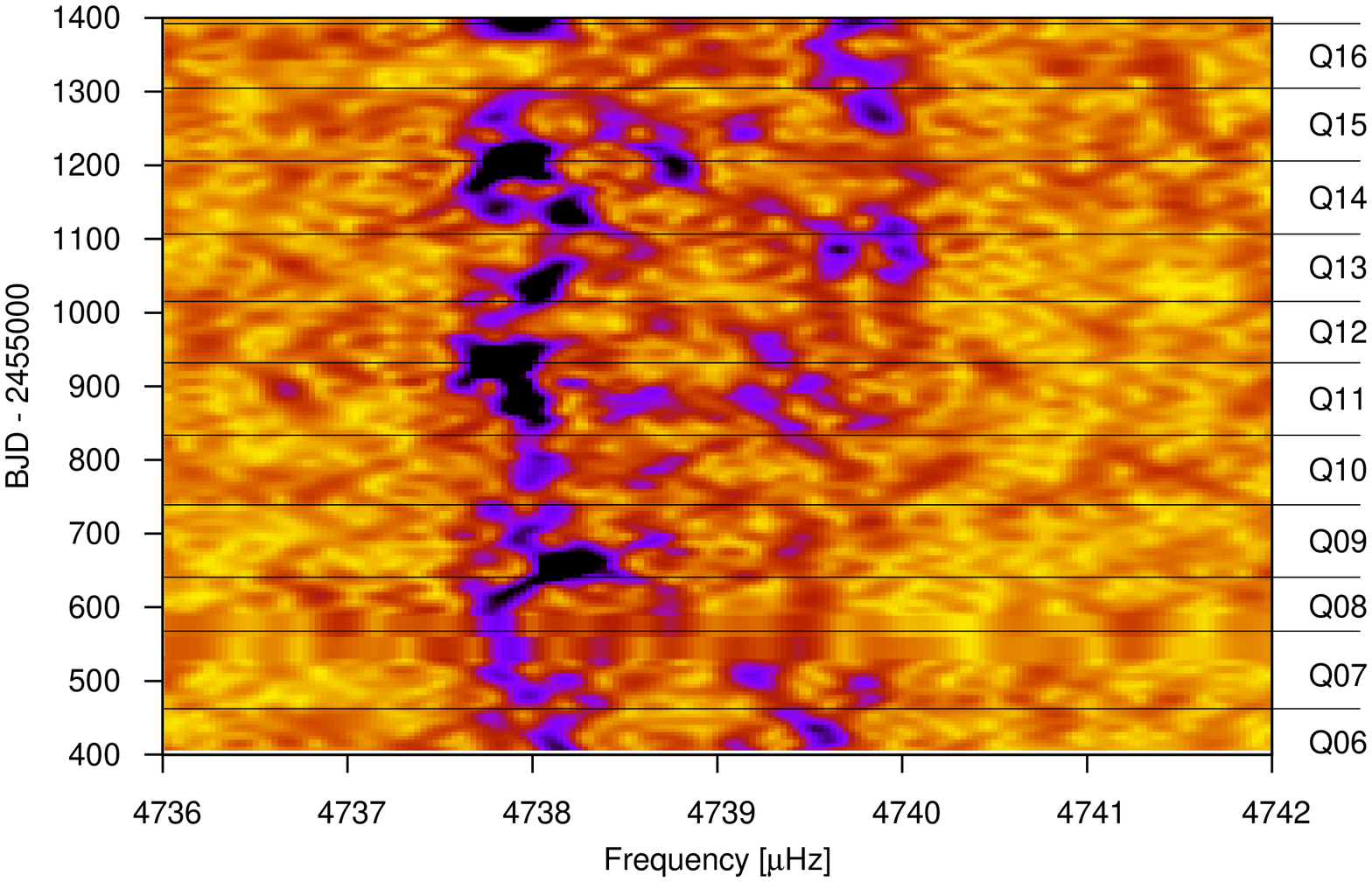,height=9cm,angle=-90}}
\centerline{\psfig{figure=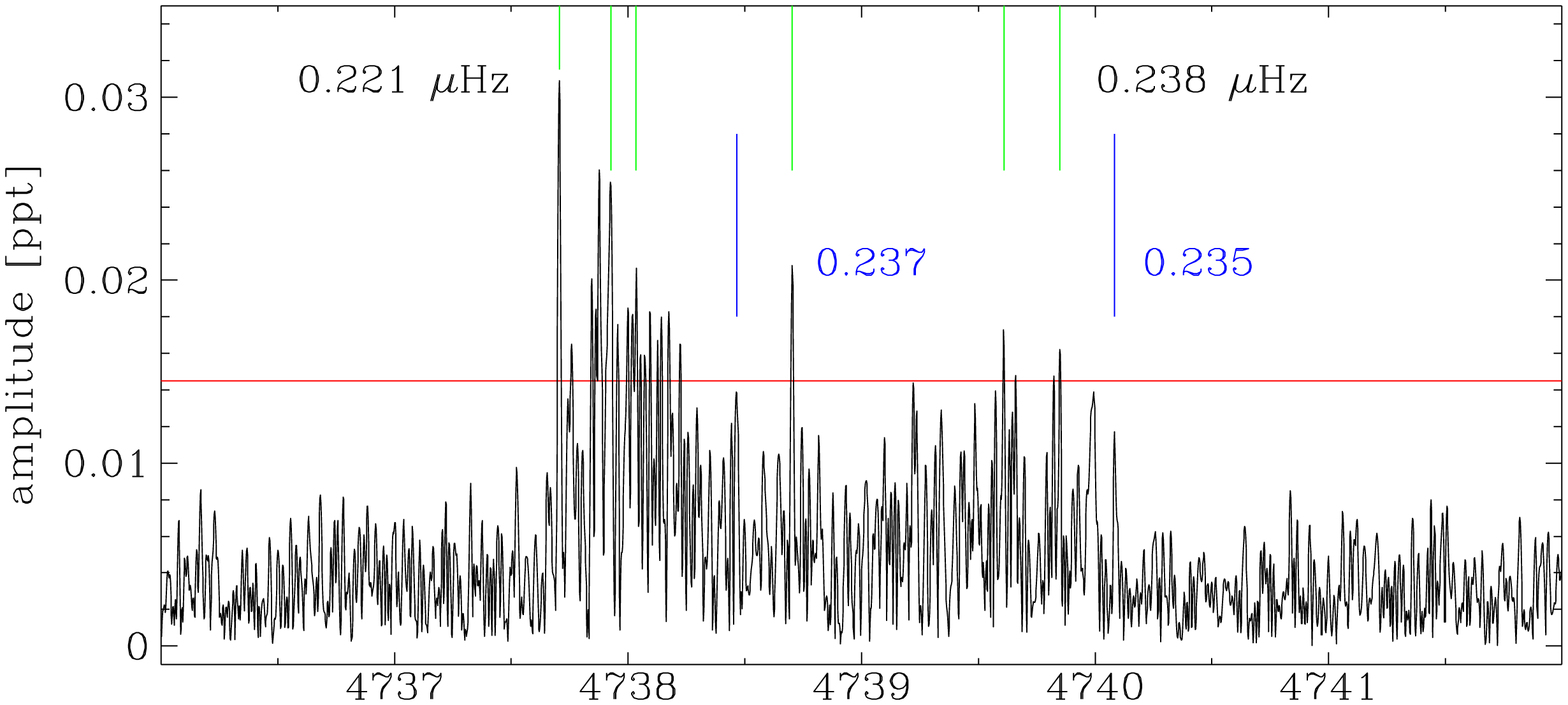,height=7.75cm,angle=-90}}
\caption[]{The $p$-mode complex at 4739\,$\mu$Hz. Top: dynamic Fourier
  { transform computed with} a running window of 50 days. The colour cuts
  range from 0 to 0.08\,ppt.  Bottom: FT of the full Q06--Q17
  \kep\ light curve, with the 6 modes that are listed in
  Table~\ref{tbl:freqlist} indicated by green vertical line segments, and
  the adopted detection threshold by the horizontal line.  The typical
  0.22-0.24\,$\mu$Hz splitting is indicated for two pairs of
  these. Two { more such} splittings, involving less significant peaks,
  are indicated in blue.
  The broad power humps in the full-length FT reflect the short
  life times of these modes.}
\label{fig:pmodes}
\end{figure}

\subsection{Radial rotation profile and $p$-modes}

So far in this section we have assumed that the star has rigid
rotation.  In case of a non-rigid radial rotation profile,
$\Omega(r)$, the multiplet splittings may be expected to vary along
with the radial order of the pulsations, as different parts of the
stellar interior are probed through pulsations of different radial
order.  Similarly, the splittings of $p$-modes reflect the rotation
frequency in the outer envelope, whereas the $g$-mode splittings are a
measure of the rotation frequency further inwards, probing all the way
in to the base of the He mantle, outside the He-burning convective
core. See e.g.\ \citet[][]{charpinet00,charpinet14} for pulsation mode
propagation as a function of stellar radius.

\begin{figure*}[t]
\centerline{\psfig{figure=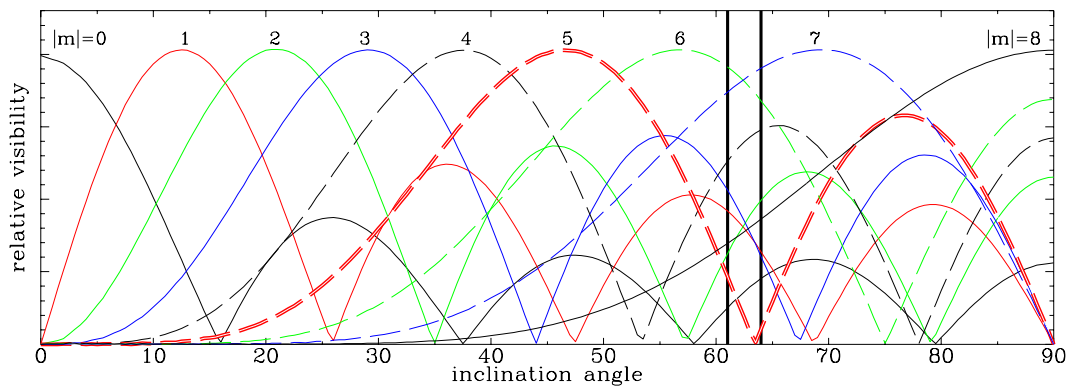,height=14cm,angle=-90}}
\caption[]{Relative visibility of $\ell$=8 modes as a function of
  inclination angle. Between $i$$\sim$61$\degr$ and $i$$\sim$64$\degr$ 
  the $m$=0 and $|m|$=5 components have low visibility, while the
  other multiplet components are visible, similar to what is
  observed (see top panels of Fig.~\ref{fig:complex}). }
\label{fig:visibilityL8}
\end{figure*}

In the typical sdB $p$-mode region with periods shorter than 15
minutes we only find 5 scattered low but significant peaks in the
region of 1900--2600\,$\mu$Hz, and several more in a complex pattern
of peaks at 4739\,$\mu$Hz.  For the modes in this complex pattern we
present a dynamic Fourier diagram in Fig.~10, for which we have used
chunks of 50 days to compute the running FTs.  The figure shows that
the modes in this pattern are short lived, with life times on the same
order as the rotation period.  Among the splittings between subsequent
peaks, as listed in Table~\ref{tbl:freqlist}, we find two that have
the usual values of 0.22--0.24\,$\mu$Hz, and we find another two such
splittings that involve two less-significant peaks (see Fig.\ 10).
With a $p$-mode value of $C_{n\,\ell}$$\sim$0, the splittings as
indicated in the figure correspond to a rotation-period range of
49--52\,d, i.e.\ a similar but slightly longer rotation period as
derived from the $g$-modes, implying a constant or slightly decreasing
rotation frequency as a function of radius.  We note, however, that
the $p$-mode frequencies are few, and are difficult to interpret, with
no observational anchor points such as clear multiplets to hold on to.

Another clue towards differential rotation may be presented by the
different values of the rotational splitting of the $\ell$=1 triplets
presented in Fig.~\ref{fig:multiplets}. These triplets are placed in a
narrow range of radial order, but their splittings differ from triplet
to triplet.  The largest difference is between the triplet at
345.7\,$\mu$Hz, with radial order $n$=9 and splitting of
0.118\,$\mu$Hz, and that at 295.0\,$\mu$Hz, with radial order $n$=11
and splitting of 0.139\,$\mu$Hz. When assuming rigid rotation, this
difference corresponds to a difference in $C_{n\,\ell}$ values of
almost 18\%, which is large for what one may expect for high-$n$
$g$-modes \citep[see e.g.][]{charpinet00}, and may in itself be
evidence for non-rigid rotation.

\subsection{Inclination angle}

Even though many multiplets are incomplete, there are some that
provide us information about the inclination angle of the star.  In
general, mode visibility is a function of inclination angle, and for
some inclination angles that are close to modal node lines, particular
modes in multiplets are not expected to be visible.

When taken all together, the $\ell$=1 triplets in
Fig.~\ref{fig:multiplets} show that the central $m$=0 components are
not as strong as the $|m|$=1 components, which points at an
inclination angle larger than 45$\degr$, but not close to 90$\degr$.

Secondly, the $\ell$=8 multiplet shown in Fig.~\ref{fig:complex} can
be interpreted as having a very low amplitude $m$=0 component at
1041.1\,$\mu$Hz, and similarly low $|m|$=5 components.  To simulate
the expected visibility of an $\ell$=8 multiplet as a function of
inclination angle we adapted the model described by
\citet{schrijvers1999}, that was originally used for modelling of
pulsational line-profile variations.  We assume that the brightness
variation along the stellar surface can be described by a spherical
harmonic with sinusoidal amplitude as a function of time.  The result
of this exercise is shown in Fig.~\ref{fig:visibilityL8}, and one can
infer that for an inclination angle close to $i$$\sim$62$\degr$ the
$m$=0 and $|m|$=5 components have low visibility, while the other
components are better visible.  In case the
observed structure reflects an $\ell$=9 multiplet instead, a similar
constraint on the inclination angle can be derived:
$i$$\sim$55$\degr$.

For an orbital inclination of $i$$\sim$62$\degr$\,, which implies
spin-orbit alignment, the orbital RV solution (Table 1) constrains the
mass of the WD companion to 0.48\,$M_{\sun}$, if one assumes a
canonical mass of 0.47\,$M_{\sun}$ for the sdB.

\section{Summary and Conclusions}

From our new low-resolution spectroscopy and the \kep\ lightcurve we
discovered that \target\ is a binary consisting of a B subdwarf and an
unseen companion, likely a white dwarf. We found a circular orbit with
$P_{\rm orb}$=14.2\,d, and a radial-velocity amplitude of 39km\,s$^{-1}$.  

The 2.88\,years of short-cadence \kep\ data of \target\ reveal Doppler
beaming at the 163 ppm level.  The light-travel delay measured by
timing of the pulsations was derived at 27\,s.  Both the observed
amplitude of the Doppler beaming and the measured light-travel
delay are consistent with the spectroscopic orbital radial-velocity
curve. The Doppler-beaming amplitude and the lack of the reflection
effect both imply an unseen compact companion.

From the high signal-to-noise average spectrum we redetermined the
atmospheric parameters of the subdwarf: \teff\,=\,27\,700\,K and
\logg\,=\,5.50\,dex (LTE values).  Nitrogen and iron have abundances
close to the solar values, while helium, carbon, oxygen and silicon
are underabundant relative to the solar mixture.

We extracted 132 significant pulsation frequencies from the
\kep\ light curve, with strong evidence for additional pulsational
content near the detection limit.  We found several $p$-modes, and
many frequencies in the $g$-mode domain, demonstrating the potential
for a seismic analysis of this star.  Many, if not all, of the
pulsation modes show amplitude variability.  For $\ell$=1 multiplets
at low frequencies, 127--170 $\mu$Hz, some modes of high amplitude
have short-lived epochs of very high amplitude, i.e. with a duration
of a few times the rotation period. Many other modes of higher $\ell$
were found to be short lived, with the shortest life times comparable
to the rotation period.

Long- and short-term amplitude variability has been observed in other
sdB stars as well, see e.g. \citet{reed07b} and \citet{ostensen14b}
who document also frequency and phase variability of sdB-star
pulsations.  These variations point at structural variability of the
pulsation cavities and/or redistribution of pulsation energy over
different modes.  The above authors suggest that phase variability may
originate in convective motions, which are somehow interacting with
the Z-bump driving mechanism.  The long time base observations of
\target\ that we present here confirm that amplitude variability is
indeed an important phenomenon in sdB-star pulsations.

We used four clear $\ell$=1 triplets to anchor the period-spacing
sequences.  We showed that many of the pulsation frequencies match
period spacings of $\ell$=1 and $\ell$=2 sequences, with no obvious
evidence for mode trapping. These period-spacing sequences will aid in
the identification process of the modes in a future seismic study of
this object. From these sequences we identify many $\ell$=1 $g$-modes
in the range of radial orders $n$$\sim$4--29, and $\ell$=2 $g$-modes
within $n$$\sim$17--61.

We find many multiplets, and many have amplitude-variable components.
The long time base of the \kep\ observations allows us to see many
multiplet components, even though they are not all excited at the same
time. Nevertheless, in spite of the extensive time coverage many
multiplets remain incomplete.

There are several frequency regions that show multiplets of
high-degree $\ell$$\ge$3 modes.  As for the low-degree multiplets in
\target, all are incomplete or have low-amplitude components.  Even
though some of these components are below the 4.5-$\sigma$ detection
limit, the patterns of equally-split frequencies are striking and make
the detection of these high-$\ell$ multiplets quite reliable.  The
most convincing of these multiplets is indicative of modes with degree
$\ell$=8, which is a novelty in observational sdB-star seismology.

Given that the surface cancellation effects in photometry are high,
these high-degree modes must be intrinsically strong in \target.  For
other types of pulsating stars with rapid rotation, high-degree modes
are typically found from time-resolved spectroscopy: the stellar
surface is resolved through rotational broadening of the absorption
lines, which allows Doppler imaging. An example of a star with a
dominant $\ell$=8 mode is the $\beta$ Cephei variable $\omega^1$ Sco
\citep{telting1998,berdyugina2003}, showing clear spectroscopic
variability, while because of its photometric stability the star was
one of the prime standard stars of the Walraven photometric system.
Although recent space missions have shown very dense pulsation spectra
in different types of stars, we are unaware of any star for which
an $\ell$=8 multiplet has been identified from photometry until
now.

From mode-visibility considerations we derive that the inclination of
the rotation axis of the sdBV in \target\ must have an intermediate
value around $\sim$60$\degr$.

From the $g$-mode multiplet splitting we derive a rotation period of
46--48\,d for \target, at the inner depths of the He mantle sampled by
these modes. The few, and difficult to interpret, $p$-mode splittings
may point at a somewhat slower rotation rate further out in the
envelope.  Additional evidence for non-rigid rotation may come from
the large difference between the splittings of the four main $\ell$=1
triplets.  Detailed modelling of the observed splittings of the modes
spanning almost 60 radial orders \citep[e.g.][]{beck2012}, has the
potential to reveal further evidence for non-rigid rotation in this
star.

As in all sdBV binaries in the \kep\ field studied so far, the
rotation is subsynchronous with respect to the orbit.  In this
respect, and this holds also for the main atmospheric and the orbital
parameters, \target\ lives in the same observable parameter space as
\object{KIC\,11558725} \citep{telting12}.

\target\ is the third sdB pulsator in the \kep\ sample with a
confirmed compact stellar-mass companion.  \target\ has the
fifth-longest orbital period of all known sdB stars with compact
companions, which is at the long end of the period range of the
$\sim$100 known sdB binaries that have periods compatible with the
common-envelope ejection scenario.

Assuming a canonical sdB mass of 0.47\,$M_{\sun}$, we derived a lower
limit for the mass of the companion of 0.40\,$M_{\sun}$.  The distance
between the two companions is $\ga$23\,$R_{\sun}$, which implies that
if the sdB is a result of a common-envelope phase the progenitor must
have been close to the maximum radius for a red giant, near the tip of
the red-giant branch.

\begin{acknowledgements}

Based on observations made with the Nordic Optical Telescope, operated
on the island of La Palma jointly by Denmark, Finland, Iceland,
Norway, and Sweden, in the Spanish Observatorio del Roque de los
Muchachos (ORM) of the Instituto de Astrofisica de Canarias, and the
William Herschel Telescope also at ORM, operated by the Isaac Newton
Group.

This paper includes data collected by the Kepler mission.  The authors
gratefully acknowledge the \kep\ team and all who have contributed to
enabling the mission.  The \kep\ data presented in this paper were
obtained from the Mikulski Archive for Space Telescopes (MAST).
Funding for the \kep\ Mission is provided by NASA's Science Mission
Directorate.

ASB gratefully acknowledge a financial support from the Polish
National Science Centre under project No.\,UMO-2011/03/D/ST9/01914.

TK acknowledges support by the Netherlands Research
School for Astronomy (NOVA).

JHT thanks the referee for the constructive comments.

The research leading to these results has received funding from the
European Community's Seventh Framework Programme FP7-SPACE-2011-1,
project number 312844 (SPACEINN).

\end{acknowledgements}

\bibliographystyle{aa}
\bibliography{sdbrefs}

\Online

\appendix

\section{Tables: measured radial velocities, fitted metal lines, and extracted
frequencies with mode identifications}

\begin{table*}[ht]
\begin{center}
\caption[]{Log of the low-resolution spectroscopy of \target\ and the RV measurements}
\label{tbl:obslog}
\begin{tabular}{ccrr@{~~}rll}
\hline\hline \noalign{\smallskip}
Mid-exposure Date & {Barycentric JD} & S/N 
& RV~~~ & RV$_{\rm err}$~ & Telescope & Observer/PI \\
   & --2450000  & & km\,s$^{-1}$ & km\,s$^{-1}$ \\
\noalign{\smallskip} \hline \noalign{\smallskip}
2010-07-27T22:33:49.8 & 5405.44261  & 85.4  &  $-$59.9 &   2.4 &  WHT  & JHT,CA \\ 
2010-07-28T01:46:03.9 & 5405.57610  & 84.0  &  $-$68.6 &   6.0 &  WHT  & JHT,CA \\ 
2011-06-01T00:09:55.7 & 5713.50851  & 62.7  &  $-$33.2 &  12.4 &  NOT  & JHT    \\ 
2011-06-01T04:32:37.3 & 5713.69094  & 50.7  &  $-$55.7 &  12.8 &  NOT  & JHT    \\ 
2011-06-07T05:38:55.6 & 5719.73717  & 18.9  &  $-$13.3 &  35.9 &  NOT  & JHT    \\ 
2011-06-20T00:47:24.7 & 5732.53504  & 48.9  &  $-$39.0 &  14.6 &  NOT  & JHT    \\ 
2011-06-20T03:04:04.2 & 5732.62995  & 42.7  &  $-$33.8 &  18.5 &  NOT  & JHT    \\ 
2011-07-23T21:41:22.0 & 5766.40619  & 61.6  &      4.6 &   9.6 &  NOT  & JHT    \\ 
2011-07-23T23:11:50.8 & 5766.46902  & 49.1  &      9.8 &  13.6 &  NOT  & JHT    \\ 
2011-07-24T00:52:21.8 & 5766.53883  & 48.3  &     17.4 &  15.3 &  NOT  & JHT    \\ 
2011-07-24T02:32:25.2 & 5766.60831  & 58.2  &     20.1 &  11.1 &  NOT  & JHT    \\ 
2011-07-24T03:52:13.3 & 5766.66373  & 52.4  &   $-$0.0 &   8.4 &  NOT  & JHT    \\ 
2013-06-02T03:40:12.2 & 6445.65459  & 76.5  &   $-$4.6 &  11.3 &  NOT  & JHT    \\ 
2013-06-03T03:32:11.5 & 6446.64906  & 60.4  &     10.2 &   6.9 &  WHT  & SM,TK  \\ 
2013-06-03T03:45:30.7 & 6446.65831  & 60.9  &     16.5 &   8.0 &  NOT  & JHT    \\ 
2013-06-04T03:30:49.1 & 6447.64814  & 67.5  &      0.1 &   3.6 &  WHT  & SM,TK  \\ 
2013-06-05T03:01:42.6 & 6448.62795  & 77.9  &   $-$7.4 &   4.9 &  WHT  & SM,TK  \\ 
2013-06-06T02:07:07.8 & 6449.59008  & 69.6  &  $-$31.2 &   9.0 &  WHT  & SM,TK  \\ 
2013-06-14T05:02:21.6 & 6457.71198  & 78.0  &  $-$16.4 &   9.2 &  NOT  & JHT    \\ 
2013-06-22T03:19:35.4 & 6465.64078  & 61.0  &  $-$59.7 &  13.6 &  NOT  & JHT    \\ 
2013-07-08T03:33:28.0 & 6481.65065  & 60.3  &  $-$69.9 &  10.7 &  NOT  & JHT    \\ 
2013-08-02T20:57:10.9 & 6507.37547  & 37.5  &  $-$41.4 &  12.8 &  NOT  & JHT    \\ 
2013-08-15T00:49:00.7 & 6519.53630  & 61.6  &   $-$8.8 &   7.7 &  NOT  & JHT    \\ 
2013-08-23T20:58:37.2 & 6528.37614  & 46.9  &  $-$12.9 &   5.4 &  NOT  & JHT    \\ 
2013-08-25T20:51:56.3 & 6530.37145  & 75.0  &     13.9 &  12.9 &  NOT  & JHT    \\ 
2013-08-26T21:00:01.3 & 6531.37704  & 75.8  &      9.7 &  10.0 &  NOT  & JHT    \\ 
2013-08-28T23:26:33.5 & 6533.47875  & 70.1  &   $-$8.9 &   7.6 &  NOT  & JHT    \\ 
2013-09-07T20:30:43.9 & 6543.35638  & 47.2  &      5.2 &   7.5 &  NOT  & JHT    \\ 
2013-09-10T21:02:03.8 & 6546.37804  & 62.9  &      7.1 &   7.4 &  NOT  & JHT    \\ 
2013-09-22T20:10:49.5 & 6558.34205  & 67.6  &      3.8 &  10.3 &  NOT  & JHT    \\ 
\noalign{\smallskip} \hline            
\end{tabular} \end{center} \end{table*}

\newpage

\begin{table}[b]
\caption[]{Fitted metal lines with equivalent widths larger than
50\,m\AA.}
\label{table:lines}
\centering
\begin{tabular}{lcr|lcr} \hline\hline \noalign{\smallskip}
Ion   & Wavelength & $W_\lambda$ & Ion   & Wavelength  & $W_\lambda$  \\
      & [\AA]      & [m\AA] & & [\AA]      & [m\AA] \\
\noalign{\smallskip} \hline \noalign{\smallskip}
He {\sc i} & 3819.603 &  123.7 & N {\sc ii} & 4227.736 & 77.1 \\
He {\sc i} & 3819.614 &  54.5  & N {\sc ii} & 4237.047 & 66.3 \\
He {\sc i} & 3888.605 &  106.1 & N {\sc ii} & 4241.755 & 59.8 \\
He {\sc i} & 3888.646 &  126.1 & N {\sc ii} & 4241.786 & 79.7 \\
He {\sc i} & 3888.649 &  150.6 & N {\sc ii} & 4432.736 & 64.8 \\
He {\sc i} & 3964.729 &  61.6  & N {\sc ii} & 4442.015 & 65.6 \\
He {\sc i} & 4026.187 &  144.1 & N {\sc ii} & 4447.030 & 91.9 \\
He {\sc i} & 4026.187 &  53.2  & N {\sc ii} & 4530.410 & 72.1 \\ 
He {\sc i} & 4026.198 &  53.2  & N {\sc ii} & 4552.522 & 64.1 \\
He {\sc i} & 4026.199 &  98.9  & N {\sc ii} & 4601.478 & 74.5 \\
He {\sc i} & 4026.358 &  73.1  & N {\sc ii} & 4607.153 & 66.6 \\
He {\sc i} & 4143.761 &  76.5  & N {\sc ii} & 4621.393 & 65.7 \\
He {\sc i} & 4387.929 &  89.0  & N {\sc ii} & 4630.539 & 114.0 \\
He {\sc i} & 4471.473 &  207.5 & N {\sc ii} & 4643.086 & 82.9 \\
He {\sc i} & 4471.473 &  102.7 & N {\sc ii} & 4678.135 & 83.7 \\
He {\sc i} & 4471.485 &  102.7 & N {\sc ii} & 4803.287 & 52.1 \\
He {\sc i} & 4471.488 &  177.1 & N {\sc ii} & 5001.134 & 93.9 \\  
He {\sc i} & 4471.682 &  127.0 & N {\sc ii} & 5001.474 & 107.9 \\   
He {\sc i} & 4713.139 &  88.3  & N {\sc ii} & 5005.150 & 117.0 \\
He {\sc i} & 4713.156 &  52.2  & N {\sc ii} & 5007.328 & 78.4 \\
He {\sc i} & 4921.931 &  188.2 & N {\sc ii} & 5045.099 & 75.3 \\
He {\sc i} & 5015.678 &  151.3 & O {\sc ii} & 4072.153 & 68.1 \\
N {\sc ii} & 3955.851 & 59.3   & O {\sc ii} & 4349.426 & 63.8 \\
N {\sc ii} & 3994.997 & 109.0  & O {\sc ii} & 4641.810 & 55.8 \\
N {\sc ii} & 4035.081 & 73.7   & O {\sc ii} & 4649.135 & 55.3 \\
N {\sc ii} & 4041.310 & 89.2   & Si {\sc iii} & 4552.622 & 66.8 \\
N {\sc ii} & 4043.532 & 82.2   & Si {\sc iii} & 4567.840 & 55.0 \\
N {\sc ii} & 4082.270 & 53.2   & Fe {\sc iii} & 4164.731 & 51.6 \\
N {\sc ii} & 4176.159 & 71.4   & Fe {\sc iii} & 5156.111 & 55.7 \\  
\noalign{\smallskip} \hline
\end{tabular}\end{table}



\begin{table*}[ht]
\begin{center}
\caption[]{Extracted pulsation frequencies. Three amplitudes are
  listed, based on multi-sine fits: the full Q06--17 amplitude, with
  errors of 2.6\,ppm , and the amplitudes A and B of the first and
  second halves of the data, with error 3.8\,ppm.  For the S/N column
  we used the error estimate of 3.2\,ppm based on the mean (residual)
  Fourier amplitude.  Frequency differences, or splittings, between
  subsequent table entries are listed when smaller than 4.5\,$\mu$Hz.
  Radial orders are listed for modes belonging to the $\ell$=1 and
  $\ell$=2 period sequences.  Vertical arrows and horizontal
  delimiters indicate the multiplet ranges, with the radial order
  listed for the multiplet component that best matches the central
  component.}
\label{tbl:freqlist}
\begin{tabular}{r@{}llllrrrrrrr}
\hline\hline \noalign{\smallskip}
\multicolumn{3}{c}{Frequency}& \multicolumn{2}{c}{Period} & \multicolumn{1}{c}{Ampl}& S/N & \multicolumn{2}{c}{Ampl A, B} &  Splitting                   & $n_{\ell=1}$ & $n_{\ell=2}$ \\
\multicolumn{3}{c}{$\mu$Hz}  & \multicolumn{2}{c}{s}      & \multicolumn{1}{c}{ppm}      &     & \multicolumn{2}{c}{ppm}            &  \multicolumn{1}{r}{$\mu$Hz} \\ 
\noalign{\smallskip} \hline \noalign{\smallskip}
%
%
%
   4.&8378   & 0.0007   &  206705       & 30      &   23 &   7.3 &   14 &   32 &        &    &     \\ %
  35.&0779   & 0.0011   &   28507.9     & 0.9     &   15 &   4.8 &   11 &   19 &        &    &     \\ %
  40.&1130   & 0.0011   &   24929.6     & 0.7     &   14 &   4.5 &   17 &   11 &  0.239 &    &     \\ %
  40.&3524   & 0.0007   &   24781.67    & 0.43    &   23 &   7.2 &   20 &   25 &        &    &     \\ %
  78.&0709   & 0.0009   &   12808.87    & 0.14    &   19 &   6.0 &   18 &   19 &  0.276 &    &     \\ %
  78.&3471   & 0.0006   &   12763.71    & 0.10    &   27 &   8.6 &   27 &   27 &  0.244 &    &     \\ %
  78.&5909   & 0.0007   &   12724.13    & 0.11    &   23 &   7.3 &   32 &   14 &        &    &     \\ \cline{12-12} %
 109.&7497   & 0.0010   &    9111.64    & 0.08    &   16 &   5.2 &   12 &   21 &  0.241 &    & \ua \\ %
 109.&9911   & 0.0008   &    9091.65    & 0.06    &   21 &   6.7 &   25 &   17 &        &    &  61 \\ \cline{12-12} %
 115.&3917   & 0.0006   &    8666.132   & 0.048   &   25 &   8.0 &   26 &   24 &  0.254 &    & \ua \\ %
 115.&64534  & 0.00042  &    8647.128   & 0.031   &   39 &  12.3 &   26 &   52 &  0.267 &    &     \\ %
 115.&91265  & 0.00012  &    8627.186   & 0.009   &  139 &  43.5 &  128 &  151 &  0.920 &    &  58 \\ \cline{12-12} %
 116.&8322   & 0.0008   &    8559.29    & 0.06    &   19 &   6.2 &   20 &   20 &  0.233 &    &     \\ %
 117.&0650   & 0.0008   &    8542.26    & 0.06    &   21 &   6.6 &   27 &   14 &  1.043 &    &     \\ %
 118.&1078   & 0.0008   &    8466.84    & 0.06    &   20 &   6.3 &   28 &   15 &  1.389 &    &     \\ \cline{12-12} %
 119.&49731  & 0.00008  &    8368.389   & 0.006   &  202 &  63.3 &  228 &  176 &  0.246 &    &  56 \\ %
 119.&74315  & 0.00009  &    8351.208   & 0.007   &  174 &  54.4 &   98 &  252 &  0.236 &    &     \\ %
 119.&97911  & 0.00011  &    8334.784   & 0.008   &  144 &  45.2 &  158 &  130 &        &    & \da \\ \cline{12-12} %
 127.&1452   & 0.0006   &    7865.024   & 0.036   &   27 &   8.7 &   27 &   27 &  4.192 & 29 &     \\ %
 131.&33672  & 0.00048  &    7614.017   & 0.028   &   33 &  10.5 &   41 &   27 &  0.196 &    &     \\ %
 131.&5328   & 0.0008   &    7602.664   & 0.047   &   20 &   6.3 &   18 &   22 &        &    &     \\ \cline{11-11} %
 145.&67445  & 0.00014  &    6864.622   & 0.007   &  116 &  36.4 &   70 &  162 &  0.138 & 25 &     \\ %
 145.&81287  & 0.00019  &    6858.105   & 0.009   &   87 &  27.3 &  102 &   63 &  0.289 & \da&     \\ \cline{11-12} %
 146.&10222  & 0.00027  &    6844.523   & 0.013   &   61 &  19.1 &   65 &   53 &  0.236 &    & \ua \\ %
 146.&33824  & 0.00046  &    6833.484   & 0.021   &   35 &  11.1 &   43 &   27 &  0.211 &    &     \\ %
 146.&5493   & 0.0008   &    6823.642   & 0.038   &   19 &   6.2 &   10 &   22 &  4.337 &    &  45 \\ \cline{12-12} %
 150.&88592  & 0.00042  &    6627.524   & 0.019   &   38 &  12.1 &   42 &   36 &  1.073 & 24 &     \\ %
 151.&9591   & 0.0007   &    6580.720   & 0.030   &   23 &   7.5 &   32 &   14 &  0.208 &    &     \\ %
 152.&1670   & 0.0007   &    6571.726   & 0.032   &   22 &   6.9 &   16 &   28 &        &    &     \\ %
 157.&23424  & 0.00049  &    6359.938   & 0.020   &   33 &  10.3 &   25 &   40 &  1.768 & 23 &     \\ \cline{12-12} %
 159.&0026   & 0.0006   &    6289.206   & 0.024   &   26 &   8.3 &   39 &   14 &  0.254 &    & \ua \\ %
 159.&2571   & 0.0006   &    6279.156   & 0.025   &   25 &   8.1 &   26 &   24 &  0.237 &    &     \\ %
 159.&4943   & 0.0010   &    6269.816   & 0.037   &   17 &   5.4 &   21 &   13 &  3.975 &    &  41 \\ \cline{12-12} %
 163.&4696   & 0.0010   &    6117.347   & 0.036   &   17 &   5.3 &   23 &   10 &  0.837 & 22 &     \\ %
 164.&3066   & 0.0006   &    6086.183   & 0.023   &   26 &   8.2 &   12 &   41 &        &    &  40 \\ \cline{11-11} %
 170.&02516  & 0.00015  &    5881.482   & 0.005   &  112 &  35.2 &   71 &  159 &  0.163 & 21 &     \\ %
 170.&18782  & 0.00011  &    5875.8610  & 0.0038  &  150 &  47.1 &  110 &  195 &  2.179 & \da&     \\ \cline{11-11} %
 172.&3668   & 0.0007   &    5801.583   & 0.023   &   23 &   7.5 &   37 &   11 &        &    &  38 \\ %
 189.&1709   & 0.0006   &    5286.226   & 0.017   &   27 &   8.5 &   20 &   34 &        &    &     \\ \cline{11-11} %
 194.&498567 & 0.000039 &    5141.4260  & 0.0010  &  420 & 131.4 &  450 &  386 &  0.125 & \ua&     \\ %
 194.&62319  & 0.00048  &    5138.134   & 0.013   &   34 &  10.7 &   49 &   22 &  0.128 & 18 &     \\ %
 194.&75097  & 0.00012  &    5134.7627  & 0.0033  &  131 &  41.2 &  169 &   93 &  1.118 & \da&     \\ \cline{11-11} \cline{12-12} %
 195.&86917  & 0.00015  &    5105.4487  & 0.0038  &  112 &  35.1 &  143 &   81 &  0.243 &    & \ua \\ %
 196.&11181  & 0.00012  &    5099.1319  & 0.0033  &  130 &  40.7 &  171 &   90 &  0.231 &    &     \\ %
 196.&3424   & 0.0007   &    5093.145   & 0.019   &   22 &   7.0 &    9 &   37 &  0.225 &    &  33 \\ %
 196.&56709  & 0.00042  &    5087.322   & 0.011   &   38 &  12.1 &   39 &   37 &  0.242 &    &     \\ %
 196.&8089   & 0.0007   &    5081.071   & 0.019   &   22 &   6.9 &    8 &   36 &  4.159 &    & \da \\ \cline{12-12} %
 200.&9686   & 0.0010   &    4975.913   & 0.024   &   16 &   5.3 &   21 &   11 &  0.504 &    &  32 \\ %
 201.&4718   & 0.0006   &    4963.475   & 0.014   &   28 &   8.9 &   26 &   31 &  2.781 &    & \da \\ \cline{11-11} \cline{12-12} %
 204.&2523   & 0.0007   &    4895.907   & 0.016   &   24 &   7.6 &   30 &   21 &  0.261 & \ua&     \\ %
 204.&51310  & 0.00007  &    4889.6623  & 0.0017  &  235 &  73.6 &  318 &  152 &        & 17 &     \\ \cline{11-11} \cline{12-12} %
 219.&99408  & 0.00027  &    4545.577   & 0.005   &   61 &  19.2 &   79 &   46 &  0.227 &    & \ua \\ %
 220.&22068  & 0.00038  &    4540.900   & 0.008   &   42 &  13.4 &   52 &   33 &  0.498 &    &  29 \\ %
 220.&7187   & 0.0007   &    4530.655   & 0.015   &   22 &   6.9 &   27 &   15 &        &    & \da \\ \cline{12-12} %
 226.&539838 & 0.000043 &    4414.2346  & 0.0008  &  376 & 117.7 &  418 &  335 &  3.064 & 15 &     \\ %
\noalign{\smallskip} \hline
\end{tabular} \end{center} \end{table*}

\addtocounter{table}{-1}
\begin{table*}[ht]
\begin{center}
\caption[]{Extracted pulsation frequencies, continued.}
\begin{tabular}{r@{}llllrrrrrrr}
\hline\hline \noalign{\smallskip}
\multicolumn{3}{c}{Frequency}& \multicolumn{2}{c}{Period} & \multicolumn{1}{c}{Ampl}& S/N & \multicolumn{2}{c}{Ampl A, B} &  Splitting                   & $n_{\ell=1}$ & $n_{\ell=2}$ \\
\multicolumn{3}{c}{$\mu$Hz}  & \multicolumn{2}{c}{s}      & \multicolumn{1}{c}{ppm}      &     & \multicolumn{2}{c}{ppm}            &  \multicolumn{1}{r}{$\mu$Hz} \\
\noalign{\smallskip} \hline \noalign{\smallskip}
 229.&6037   & 0.0008   &    4355.330   & 0.014   &   21 &   6.8 &   25 &   18 &  4.409 &    &     \\ %
 234.&0129   & 0.0011  &    4273.268    & 0.020   &   14 &   4.7 &   14 &   14 &  3.084 &    &     \\ %
 237.&0972   & 0.0013  &    4217.680    & 0.023   &   12 &   4.0 &   13 &   11 &  0.713 &    &     \\ %
 237.&8106   & 0.0013  &    4205.028    & 0.023   &   12 &   4.0 &   19 &    5 &  0.238 &    &     \\ %
 238.&0490   & 0.0008  &    4200.815    & 0.015   &   19 &   6.1 &   22 &   18 &  2.894 &    &     \\ \cline{11-11} %
 240.&9432   & 0.0008  &    4150.357    & 0.015   &   19 &   6.1 &   21 &   19 &  0.124 & \ua&     \\ %
 241.&0667   & 0.0006  &    4148.230    & 0.010   &   28 &   9.0 &   44 &   16 &  0.115 & 14 &     \\ %
 241.&18186  & 0.00024 &    4146.2489   & 0.0041  &   68 &  21.5 &   69 &   67 &  1.841 & \da&     \\ \cline{11-11} \cline{12-12} 
 243.&0232   & 0.0012  &    4114.833    & 0.020   &   14 &   4.4 &   23 &    0 &  0.296 &    & \ua \\ %
 243.&3197   & 0.0009  &    4109.820    & 0.016   &   17 &   5.5 &   23 &   14 &        &    &  26 \\ \cline{12-12} 
 250.&9339   & 0.0010  &    3985.113    & 0.017   &   15 &   4.9 &   19 &   11 &  0.660 &    &     \\ \cline{12-12} 
 251.&5939   & 0.0012  &    3974.660    & 0.018   &   14 &   4.4 &   13 &   15 &  0.328 &    & \ua \\ %
 251.&9216   & 0.0011  &    3969.489    & 0.017   &   15 &   4.8 &   12 &   18 &        &    &  25 \\ \cline{12-12} 
 261.&32394  & 0.00043 &    3826.668    & 0.006   &   38 &  11.9 &   52 &   24 &  0.233 &    & \ua \\ %
 261.&55671  & 0.00041 &    3823.263    & 0.006   &   39 &  12.5 &   57 &   25 &  0.262 &    &     \\ %
 261.&8184   & 0.0006  &    3819.441    & 0.009   &   26 &   8.3 &   29 &   22 &  0.244 &    &  24 \\ %
 262.&0623   & 0.0008  &    3815.887    & 0.011   &   21 &   6.6 &   19 &   24 &  0.231 &    &     \\ %
 262.&2929   & 0.0009  &    3812.532    & 0.013   &   18 &   5.7 &   21 &   15 &        &    & \da \\ \cline{11-11} \cline{12-12} 
 275.&5585   & 0.0005  &    3628.993    & 0.007   &   30 &   9.4 &   26 &   33 &  0.263 & 12 &     \\ %
 275.&8218   & 0.0005  &    3625.530    & 0.007   &   30 &   9.5 &   25 &   34 &        & \da&     \\ \cline{11-11} \cline{12-12} 
 282.&6156   & 0.0009  &    3538.375    & 0.012   &   17 &   5.5 &   14 &   20 &  0.071 &    & \ua \\ %
 282.&6866   & 0.0008  &    3537.486    & 0.010   &   20 &   6.3 &   15 &   24 &  0.169 &    &  22 \\ %
 282.&8558   & 0.0007  &    3535.371    & 0.008   &   25 &   7.8 &   25 &   26 &        &    & \da \\ \cline{11-11} \cline{12-12} 
 294.&83926  & 0.00012 &    3391.6785   & 0.0014  &  135 &  42.3 &  142 &  127 &  0.138 & \ua&     \\ %
 294.&97727  & 0.00027 &    3390.0918   & 0.0031  &   60 &  19.0 &   55 &   66 &  0.139 & 11 &     \\ %
 295.&11652  & 0.00007 &    3388.4921   & 0.0008  &  224 &  70.2 &  232 &  216 &  2.027 & \da&     \\ \cline{11-11} %
 297.&1435   & 0.0008  &    3365.378    & 0.009   &   21 &   6.7 &   26 &   16 &        &    &     \\ \cline{12-12} 
 307.&2303   & 0.0009  &    3254.887    & 0.010   &   17 &   5.6 &   19 &   16 &  0.238 &    & \ua \\ %
 307.&4684   & 0.0008  &    3252.366    & 0.009   &   19 &   6.2 &   17 &   22 &  0.443 &    &  20 \\ %
 307.&9111   & 0.0007  &    3247.691    & 0.007   &   24 &   7.7 &   37 &   12 &  0.217 &    &     \\ %
 308.&1280   & 0.0007  &    3245.405    & 0.007   &   23 &   7.3 &   23 &   23 &        &    & \da \\ \cline{11-11} \cline{12-12} %
 345.&5599   & 0.0006  &    2893.8534   & 0.0049  &   27 &   8.7 &   25 &   29 &  0.117 & \ua&     \\ %
 345.&67714  & 0.00022 &    2892.8728   & 0.0019  &   73 &  22.8 &   68 &   77 &  0.118 &  9 &     \\ %
 345.&79480  & 0.00011 &    2891.8885   & 0.0009  &  146 &  45.7 &  149 &  142 &        & \da&     \\ \cline{11-11} \cline{12-12} %
 352.&9717   & 0.0006  &    2833.0881   & 0.0044  &   29 &   9.2 &   24 &   34 &  0.626 &    & \ua \\ %
 353.&5973   & 0.0008  &    2828.076    & 0.006   &   21 &   6.7 &   24 &   18 &  0.188 &    &  17 \\ %
 353.&7854   & 0.0008  &    2826.572    & 0.007   &   19 &   6.1 &   27 &   11 &        &    & \da \\ \cline{12-12} %
 397.&0302   & 0.0009  &    2518.700    & 0.005   &   19 &   5.9 &   20 &   17 &  0.205 &    &     \\ %
 397.&2352   & 0.0011  &    2517.401    & 0.007   &   15 &   4.7 &   16 &   14 &        &    &     \\ %
 416.&21056  & 0.00042 &    2402.6300   & 0.0024  &   39 &  12.2 &   42 &   35 &        &  7 &     \\ %
 521.&9249   & 0.0007  &    1915.9846   & 0.0027  &   21 &   6.9 &   24 &   18 &        &  5 &     \\ %
 595.&73729  & 0.00040 &    1678.5923   & 0.0011  &   40 &  12.6 &   43 &   37 &        &  4 &     \\ %
 640.&1668   & 0.0006  &    1562.0928   & 0.0016  &   25 &   8.0 &   21 &   29 &  0.493 &    &     \\ %
 640.&6593   & 0.0010  &    1560.8919   & 0.0024  &   16 &   5.1 &   25 &    8 &  0.493 &    &     \\ %
 641.&1522   & 0.0009  &    1559.6922   & 0.0021  &   18 &   5.9 &   22 &   15 &  0.505 &    &     \\ %
 641.&6571   & 0.0012  &    1558.4648   & 0.0029  &   13 &   4.3 &   16 &   11 &        &    &     \\ %
 761.&9312   & 0.0007  &    1312.4544   & 0.0011  &   24 &   7.7 &   27 &   21 &  0.349 &    &     \\ %
 762.&2799   & 0.0012  &    1311.8541   & 0.0020  &   13 &   4.4 &   18 &    9 &        &    &     \\ %
 773.&8073   & 0.0010  &    1292.3115   & 0.0017  &   16 &   5.1 &   25 &    7 &  0.961 &    &     \\ %
 774.&7687   & 0.0011  &    1290.7077   & 0.0019  &   14 &   4.5 &   11 &   17 &  0.201 &    &     \\ %
 774.&9700   & 0.0009  &    1290.3725   & 0.0015  &   18 &   5.8 &   21 &   16 &        &    &     \\ %
 802.&3472   & 0.0006  &    1246.3433   & 0.0010  &   26 &   8.2 &   30 &   22 &  0.235 &    &     \\ %
 802.&5823   & 0.0012  &    1245.9781   & 0.0018  &   13 &   4.3 &   13 &   14 &  0.465 &    &     \\ %
 803.&0469   & 0.0009  &    1245.2573   & 0.0015  &   17 &   5.4 &   13 &   21 &  0.507 &    &     \\ %
 803.&5542   & 0.0011  &    1244.4712   & 0.0017  &   14 &   4.5 &    8 &   20 &  0.512 &    &     \\ %
 804.&0662   & 0.0012  &    1243.6787   & 0.0018  &   13 &   4.4 &   19 &    8 &  0.729 &    &     \\ %
 804.&7949   & 0.0011  &    1242.5525   & 0.0018  &   14 &   4.5 &   20 &    8 &        &    &     \\ %
\noalign{\smallskip} \hline
\end{tabular} \end{center} \end{table*}
            
\addtocounter{table}{-1}
\begin{table*}[ht]
\begin{center}
\caption[]{Extracted pulsation frequencies, continued.}
\begin{tabular}{r@{}llllrrrrrrr}
\hline\hline \noalign{\smallskip}
\multicolumn{3}{c}{Frequency}& \multicolumn{2}{c}{Period} & \multicolumn{1}{c}{Ampl}& S/N & \multicolumn{2}{c}{Ampl A, B} &  Splitting                   & $n_{\ell=1}$ & $n_{\ell=2}$ \\
\multicolumn{3}{c}{$\mu$Hz}  & \multicolumn{2}{c}{s}      & \multicolumn{1}{c}{ppm}      &     & \multicolumn{2}{c}{ppm}            &  \multicolumn{1}{r}{$\mu$Hz} \\
\noalign{\smallskip} \hline \noalign{\smallskip}
 956.&5266   &0.0012 &    1045.449181 &0.0013   &  13 &   4.2 &   17 &    9 &  1.149 &    &     \\ %
 957.&6754   &0.0012 &    1044.195166 &0.0013   &  13 &   4.1 &   13 &   13 &  3.927 &    &     \\ %
 961.&6021   &0.0007 &    1039.931223 &0.0008   &  21 &   6.8 &   27 &   17 &  2.408 &    &     \\ %
 964.&0104   &0.0010 &    1037.333208 &0.0010   &  17 &   5.4 &   23 &   10 &        &    &     \\ %
1002.&6649   &0.0012 &     997.342223 &0.0011   &  14 &   4.4 &   15 &   12 &  0.479 &    &     \\ %
1003.&1440   &0.0015 &     996.865830 &0.0015   &  10 &   3.4 &   11 &    9 &  0.241 &    &     \\ %
1003.&3846   &0.0012 &     996.626810 &0.0012   &  13 &   4.3 &   17 &   10 &  0.266 &    &     \\ %
1003.&6503   &0.0011 &     996.362973 &0.0011   &  15 &   4.7 &   16 &   14 &  0.245 &    &     \\ %
1003.&8955   &0.0012 &     996.119590 &0.0012   &  14 &   4.4 &   12 &   16 &  0.462 &    &     \\ %
1004.&3573   &0.0008 &     995.661598 &0.0008   &  20 &   6.4 &   25 &   15 &  0.223 &    &     \\ %
1004.&5804   &0.0008 &     995.440511 &0.0008   &  20 &   6.5 &   24 &   16 &        &    &     \\ %
1039.&4329   &0.0007 &     962.063094 &0.0007   &  22 &   7.0 &   21 &   23 &  0.248 &    &     \\ %
1039.&6813   &0.0010 &     961.833247 &0.0010   &  15 &   5.0 &   13 &   18 &  0.472 &    &     \\ %
1040.&1534   &0.0010 &     961.396636 &0.0009   &  16 &   5.3 &   13 &   20 &  0.269 &    &     \\ %
1040.&4225   &0.0010 &     961.147952 &0.0009   &  16 &   5.2 &    9 &   23 &  0.229 &    &     \\ %
1040.&6512   &0.0014 &     960.936735 &0.0013   &  11 &   3.7 &   10 &   12 &  0.251 &    &     \\ %
1040.&9019   &0.0006 &     960.705371 &0.0005   &  28 &   9.0 &   25 &   32 &  0.472 &    &     \\ %
1041.&3738   &0.0005 &     960.269967 &0.00048  &  31 &   9.9 &   19 &   44 &  0.244 &    &     \\ %
1041.&6181   &0.0007 &     960.044737 &0.0006   &  24 &   7.7 &   29 &   19 &  0.223 &    &     \\ %
1041.&8412   &0.0007 &     959.839199 &0.0006   &  24 &   7.7 &   27 &   20 &  0.273 &    &     \\ %
1042.&11420  &0.00042&     959.587735 &0.00039  &  38 &  12.1 &   55 &   20 &  0.476 &    &     \\ %
1042.&59031  &0.00047&     959.149522 &0.00043  &  35 &  11.0 &   31 &   38 &  0.274 &    &     \\ %
1042.&8648   &0.0006 &     958.897105 &0.0005   &  28 &   8.9 &   25 &   32 &  0.201 &    &     \\ %
1043.&0654   &0.0007 &     958.712639 &0.0007   &  21 &   6.9 &   10 &   33 &        &    &     \\ %
1916.&3004   &0.0006 &     521.838850 &0.00017  &  27 &   8.4 &   21 &   32 &        &    &     \\ %
1925.&6155   &0.0006 &     519.314482 &0.00017  &  26 &   8.3 &   25 &   28 &        &    &     \\ %
2004.&2215   &0.0009 &     498.946840 &0.00022  &  18 &   5.7 &   13 &   26 &        &    &     \\ %
2113.&1778   &0.0009 &     473.220938 &0.00020  &  18 &   5.7 &   19 &   16 &        &    &     \\ %
2543.&7726   &0.0014 &     393.116898 &0.00021  &  12 &   3.8 &   13 &   10 &  0.343 &    &     \\ %
2544.&1159   &0.0009 &     393.063859 &0.00014  &  18 &   5.8 &   18 &   18 &        &    &     \\ %
4737.&7062   &0.0005 &     211.072607 &0.000023 &  31 &  10.0 &   26 &   36 &  0.221 &    &     \\ %
4737.&9277   &0.0007 &     211.062740 &0.000030 &  24 &   7.6 &   13 &   33 &  0.106 &    &     \\ %
4738.&0338   &0.0008 &     211.058012 &0.000034 &  21 &   6.7 &   21 &   19 &  0.669 &    &     \\ %
4738.&7024   &0.0008 &     211.028233 &0.000035 &  21 &   6.6 &   18 &   23 &  0.907 &    &     \\ %
4739.&6097   &0.0009 &     210.987837 &0.000040 &  18 &   5.6 &    9 &   26 &  0.238 &    &     \\ %
4739.&8480   &0.0010 &     210.977229 &0.000046 &  15 &   5.0 &   12 &   17 &        &    &     \\ %
\noalign{\smallskip} \hline                                                                       
\end{tabular} \end{center} \end{table*}

\end{document}